\documentclass[12pt,a4paper, captions=figureheading]{scrartcl}
\usepackage[utf8]{inputenc}
\usepackage[T1]{fontenc}
\usepackage{lmodern}
\usepackage[affil-it]{authblk}
\usepackage{pdfpages}
\setkomafont{disposition}{\normalcolor\bfseries\rmfamily}

\usepackage[style=authoryear, citetracker=true, maxcitenames=2, hyperref=true, url=true, isbn=false, doi=true, natbib=true, backend=bibtex,bibstyle=authoryear, maxbibnames=30, sorting=nyt,sortcites, uniquename=full, uniquelist=false]{biblatex}
\usepackage{csquotes, color}
\usepackage[titletoc,toc,title]{appendix}
\usepackage[colorlinks=true, allcolors=blue]{hyperref}
\usepackage{setspace, float, placeins, verbatim, subcaption, amsmath}
\usepackage{pdflscape}
\usepackage{gensymb}
\usepackage{subcaption}
\usepackage{tocloft}
\makeatletter
\renewcommand{\numberline}[1]{{\@cftbsnum #1\@cftasnum~}\@cftasnumb}
\makeatother
\setkomafont{caption}{\small}
\setkomafont{captionlabel}{\usekomafont{caption}}

\usepackage{graphicx}
\usepackage[flushleft]{threeparttable}
\usepackage{tabularx}
\usepackage{booktabs, adjustbox}

\newenvironment{fignote}{\begin{quote}\scriptsize}{\end{quote}} 
\usepackage{pbox} 

\usepackage{xcolor}

\newcommand{\red}[1]{\textcolor{red}{#1}}

\usepackage{threeparttable}
\usepackage{rotating}
\usepackage{lscape}
\usepackage{soul} 
\usepackage{longtable}
\usepackage{tabu}
\usepackage[normalem]{ulem}
\usepackage[en-US]{datetime2} 
\DTMlangsetup{showdayofmonth=false}

\begin{document}
\vspace{-3cm}
\title{Technology and jobs: A systematic literature review}

\author{Kerstin H\"otte$^{1,2}$, Melline Somers$^{3}$, Angelos Theodorakopoulos$^{1}$\footnote{
The authors acknowledge support from the European Union Horizon 2020 Research and Innovation programme under grant agreement No 822330 TECHNEQUALITY. \linebreak 
The authors thankfully acknowledge Max Grigutsch for his excellent research assistance during the implementation of the systematic search strategy, Mark Levels and Pantelis Koutroumpis for insightful discussions and support. Special thanks are extended to Michael A. Freeman and Didier Fouarge for detailed feedback and insightful suggestions. The authors would also like to thank Elizabeth Champion for her support in reviewing the manuscript. \linebreak 
Contact information: kerstin.hotte@oxfordmartin.ox.ac.uk; melline.somers@maastrichtuniversity.nl; angelos.theodorakopoulos@oxfordmartin.ox.ac.uk
}
\vspace{0.25cm} \linebreak
\footnotesize{$^{1}$ Oxford Martin School, University of Oxford, 34 Broad Street, Oxford OX1 3BD, UK}\linebreak
\footnotesize{$^{2}$ Institute for New Economic Thinking, University of Oxford, Manor Road, Oxford OX1 3UQ, UK}\linebreak
\footnotesize{$^{3}$ Research Centre for Education and the Labour Market, Maastricht University, Tongersestraat 49, 6211 LM, NL}
}
\vspace{-2cm}
\date{\large{\today}}
\vspace{-1cm}
\maketitle
\vspace{-1.1cm}
\begin{abstract}
    
    Does technological change destroy or create jobs? 
    New technologies may replace human workers, but can simultaneously create 
    jobs if workers are needed to use these technologies or if new economic activities emerge. Furthermore, technology-driven productivity growth may increase disposable income, stimulating a demand-induced expansion of 
    employment.
    To synthesize the existing knowledge on this question, we systematically review the empirical literature on the past four decades of technological change and its impact on employment, distinguishing between five broad technology categories (ICT, Robots, Innovation, TFP-style, Other).
    Overall, we find across studies that the labor-displacing effect of technology appears to be more than offset by compensating mechanisms that create or reinstate labor. This holds for most types of technology, suggesting that previous anxieties over widespread technology-driven unemployment lack an empirical base, at least so far.
    Nevertheless, low-skill, production, and manufacturing workers have been adversely affected by technological change, and effective up- and reskilling strategies should remain at the forefront of policy making along with targeted social support systems.

\end{abstract}

\vspace{0.1cm}

\noindent\textbf{\small JEL Classification Codes:}{\small {} E24, J21, O3}\\
\noindent\textbf{\small Keywords:}{\small {} Technological Change, Labor, Literature Review
\thispagestyle{empty} \setcounter{page}{0}\newpage{}\setcounter{page}{1} }{\small \par}

\newpage
\section{Introduction}

The fear that technological advancement will make human labor obsolete is not new 
\citep{mokyr2015history}. During the first Industrial Revolution, the adoption of power looms and mechanical knitting frames gave rise to the Luddite movement protesting against this technology by destroying textile machinery out of fear of job losses and skill obsolescence. The idea that technology can render workers redundant, at least in the short run, has also been supported by influential economists like Karl Marx and David Ricardo in the nineteenth century \citep{marx1988economic, ricardo}. Others, such as Thomas Mortimer, believed that machines could displace labor more permanently \citep{mortimer1772elements}. In the same spirit, more recent concerns about massive job losses \citep{smith2017automation} stem from improved computing power, decreasing costs, and advances in machine learning and robotics \citep{brynjolfsson2014second}. 
The current debate on the labor market impact of technological change has also been fed by a number of influential studies predicting striking job losses in occupations that are susceptible to automation \citep{arntz2017, frey2017future, nedelkoska2018automation}. 
However, this view is controversial and has been questioned in recent research highlighting the positive employment effects of technology \citep{aghion2022effects, economist2022robots}. 

At the aggregate level, there is little evidence that technological change has led to widespread unemployment over the centuries. But technology-induced job losses can be significant. Many technologies are designed to save labor by replacing human workers by machinery. Think of tractors substituting manual labor, assembly lines replacing human handiwork and computer assisted technologies that increasingly replace workers in performing explicit and codifiable tasks \citep{nordhaus2007two}. 

However, economic theory points out that several compensating mechanisms can counterbalance the initial labor-saving impact of new technologies \citep[e.g.][]{vivarelli2014innovation, acemoglu2019automation,baldwin2021jobs}. First, technological change can increase the demand for labor by creating new tasks and jobs that are directly associated with the new technology.
Furthermore, if automation technology increases productivity it releases production resources which can raise the demand for labor in non-automated tasks within the same firm or industry.
Second, technology can also increase the demand for labor through the demand side when new technologies boost productivity growth leading to lower production costs and consumer prices. Moreover, new technologies can raise the marginal product of labor and capital resulting in higher wages and returns to capital. These two effects contribute to a rise in real income. If demand is sufficiently elastic and positively responds to increases in income and decreases in prices, technologies may lead to a demand-induced expansion of output \citep{bessen2020autom}.

In this paper, we referred to these channels through which technology affects the demand for labor as (1) replacement, (2) reinstatement, and (3) real income effect, and analyzed the empirical basis for each effect through a systematic literature search. We systematically identified and reviewed 127 studies published between 1988-2021 that provide evidence on technological change in industrialized economies in the post-1980s. To uncover potential underlying heterogeneity, we pinned down five broad technology groups that are predominantly studied in the literature distinguishing between (1) information and communication technology (ICT); (2) robot-diffusion; (3) innovation surveys; (4) productivity; and (5) a residual category that contains various alternative indicators. 

Overall, we found that the number of studies supporting the labor replacement effect is more than offset by the number of studies supporting the labor-creating reinstatement and real income effect. This observation is also supported by the studies analyzing the net employment effect of technological change suggesting the net impact of technology on labor to be rather positive than negative. 

The findings for the five distinct technology categories show broadly similar patterns, but with some subtle differences. For ICT, we found no evidence that the replacement effect dominates the reinstatement and real income effect combined. However, our results show that the reinstated jobs qualitatively differ from the jobs replaced. The diffusion of ICT mostly had positive employment implications for high-skill workers, non-routine labor, and service jobs. 
Also for robot-diffusion we found that the labor-saving impact is generally offset by robot-driven reinstatement of labor. In contrast to the ICT studies, robot studies remain silent on the complementarity between robots and human labor when performing tasks. Hence, the labor-creating effect of robots is most likely related to the production and maintenance of this type of technology. 

For studies that rely on innovation as a measure of technology, the employment impact depends on the type of innovation. While product innovation is mostly labor-creating, the evidence on the employment impact of process innovation is somewhat mixed. 
For the fourth category of studies relying on productivity as technology measure, we found a roughly equal balance between the number of studies suggesting support for the replacement and the two labor-creating mechanisms. The employment effects have been mostly favorable for non-production workers, high-skill labor, and service jobs. These studies support the idea that technological change leads to structural change with a reallocation of economic activity down the supply chain from more primary towards increasingly processed sectors and services \citep{kruger2008productivity, syrquin1988patterns}.
However, the net employment effects observed by these studies are rather negative than positive. 

Lastly, the findings from studies that rely on other/indirect measures of technology indicate that the labor replacing effect is offset by the labor-creating effect. The employment effects have been mostly positive for non-production labor, yet some studies also found positive employment effects for low-skilled workers, particularly in service jobs.

Overall, although we find larger support for the labor-creating effects of technological change, we are careful in concluding that technology has a net positive effect on employment. However, we do safely conclude that the labor replacing effect of technology is typically more than offset by a range of compensating mechanisms suggesting that the widespread anxiety over technology-driven unemployment lacks its empirical base. 

Our study is not the first to review the existing evidence on the effect of technology on employment, but our systematic approach contributes to the literature in several ways. First, most reviews are narrative which may be subject to the authors’ bias \citep{brown2002tech, bruckner2017frontier, calvino2018innovation, goos2018tech, mondolo2021composite, vivarelli2014innovation, aghion2022effects}. Second, earlier studies that provide systematic reviews of the literature focused on shorter time periods, i.e. 1980-2013 in \citet{ugur2018technological}, 2005-2017 in \citet{ballister2018}, 2000-2021 in \citet{perezacre2021participation}, while our review covers thirty-three years (1988-2021) which allows us to capture a more complete picture of the effects for each of the technologies examined. Third, in contrast to \citet{ugur2017litreview} who synthesized the evidence on less developed countries, we focused our review on developed economies aiming to capture the impact of technological change at the frontier. Fourth, while most earlier reviews restricted their analysis to specific types of technology, we analyzed how employment effects differ across five alternative measures of technological change. Fifth, to be as inclusive as possible, we did not limit our analysis to specific measurements of employment. Although evidence from different models (e.g. derived labor demand, skill/wage share, and decomposition analyses) do not necessarily yield comparable estimates, they are informative about the direction of the employment effect.

This analysis provides an empirical basis for the political and scientific debate on technology and labor, whereby the anxiety of a jobless future appears to be overstated given the poor empirical basis for this claim. Nevertheless, our results also suggest that low-skill, production, and manufacturing workers have been adversely affected by technological change. Hence, effective up- and reskilling strategies should remain at the forefront of policy making. Many occupations in the lower end but also in the middle part of the skill distribution will continue to evolve and demand a changing set of skills due to technological progress. Hence, the employment perspectives of relatively vulnerable groups can be significantly improved by investing in the right set of skills. Nonetheless, some workers who experience job loss might not be able to engage in upskilling or make the transition to new jobs. For these groups, targeted social support systems will be important.

This paper is structured as follows. Section \ref{sec:conceptual_framework} presents the conceptual framework and discusses the mechanisms through which technological change can increase or decrease the demand for labor. In Section \ref{sec:methods}, we discuss our methodology and Section \ref{sec:results} presents the results. Section \ref{sec:discussion} discusses the findings and Section \ref{sec:conclusion} concludes.

\section{Conceptual framework}
\label{sec:conceptual_framework}
We apply a three-stage framework to disentangle the interplay between technology and employment. Along these stages, we review the current state of empirical knowledge. 

We rely on a generic understanding of technology as the capability to transform a given set of inputs into outputs. Technological change happens when the quantity and/or quality of inputs or output change \citep{saviotti2013co, ruttan1959usher}. For example, new technologies may make production processes more efficient, enabling firms to produce the same goods with less labor or material inputs. It can also be reflected in the output when technologies enable firms to bring new products to the market. 

How does this interact with labor? Here, we focus on three key mechanisms which are incrementally more indirect. To illustrate these mechanisms, we introduce a stylized model with a generic production function:
\begin{equation}
    Q = A^Q f(A^L L, A^X X)
\end{equation}
where $Q$ is output, $L$ is the amount of labor used along with other inputs $X$ to produce a quality-adjusted output $Q$. Other inputs can be capital goods, material and intermediate inputs, or different forms of labor (e.g. different occupations or differently skilled workers). Note that the level of output is quality-adjusted and it can increase either through a higher number of units or a higher quality of a given unit. The parameters $A^L$, $A^X$, and $A^Q$ represent the production technology. The production function is non-decreasing in its arguments $A^L$, $L$, $A^X$, and $X$, whereby a higher level of technology or production inputs leads to an increase in output $Q$. Technological change may enter in different forms by changing $A^L$, $A^X$, and/or $A^Q$.

\subsection{The replacement effect}
The most direct impact of technology on employment is the so-called ``replacement effect''. This effect occurs when the adoption of a new technology enables a firm to reduce labor inputs for a given quantity of output. 

In the stylized model above, pure replacement happens if $A^L$ increases and $Q$ is constant, i.e. $dQ = 0$. This means that less labor is used but everything else remains equal. However, not every type of technological change leads to an increase in $A^L$, and even if this is the case, it only replaces labor if output $Q$ does not expand sufficiently. 

Other forms of technological change can lead to an increase in $A^X$ which means that the same amount of output can be produced with lower input requirements $X$. 
Technological change may also lead to an increase in $A^Q$ which increases the level of output $Q$ while keeping the input requirements constant. Examples are efficiency improvements reflected in a higher total factor productivity (TFP). Also product innovations such as the introduction of a new design can be captured by $A^{Q}$, if they enable firms to bring new and better products to the market while not changing their input requirements. 

Empirically, it is challenging to measure whether or not technological change is labor replacing. In particular, this effect is likely heterogeneous across industries and occupations, and often difficult to draw causal inference at a sufficiently granular level, especially when new jobs are created at the same time. 
For example, the introduction of a product innovation may coincide with changing input requirements reflected in the amount and type of labor. It may also be that labor-saving technological change does not necessarily lead to layoffs, but those employees that are no longer required to produce $Q$ find other useful tasks within the firm. 

In this research, we try to find out whether technological change has been labor-replacing in the post 1980s, i.e. whether technological change reflected by $A^L$ or $A^Q$ had a negative impact on the demand for labor. We use various indicators that allow us to draw conclusions about the existence of the replacement effect. At different levels of aggregation, changes in employment constitute our key indicator. Empirical support for the replacement effect exists if we observe a technology-induced decrease in employment in those firms, industries, and countries where the technology is adopted. Measures of employment include the employment rate, number of workers, and hours worked. The labor share of income is also indicative but not sufficient to provide evidence for the labor-saving impact of technology. Moreover, a number of studies at different aggregation levels examine changes in the relative employment of different occupational groups which we also consider as indicative for the replacement effect, such as a technology-induced increase in the ratio of high- over low-skill labor use. Again, this is only an indicative but not sufficient piece of evidence.\footnote{For example, technology-induced changes in the relative demand for low-skill labor at the aggregate level do not necessarily mean that labor was replaced. It may be that technological change enabled the emergence of a new industry that uses skilled labor differently. This may induce a shift in the relative demand, but in absolute terms not a single worker was replaced by machinery.}
We also interpret micro (worker or firm) level studies that assess the relationship between the type of tasks performed by workers and the likelihood of being displaced as indicative of the replacement effect, as some tasks may be more susceptible to automation. 
Another indicator of the replacement effect includes changes in the elasticity of substitution of labor and other inputs $X$. Technological change may alter this elasticity. A technology-induced increase in the elasticity indicates that the technological possibilities to replace labor by other inputs have improved. We also interpret this as supporting evidence.

\subsection{The reinstatement effect}
The reinstatement effect is the next indirect effect of technological change. It occurs if the adoption of a new technology induces the creation of new jobs that are directly linked to the new technology, regardless of whether or not technological change happens via $A^L$, $A^X$, or $A^Q$. 
The reinstatement effect is often associated with an increase in $Q$, otherwise technological change would be only input-saving even though the effects may be heterogeneous across different groups of employees. 
The creation of new jobs may be driven by different mechanisms that are empirically difficult to disentangle: Workers performing tasks that cannot be automated may experience a boost in productivity which increases the demand for these jobs.  Furthermore, new jobs may be created if technology enables new fields of economic activity.  

For example, an input-saving ($A^L$ or $A^X$) technology may induce the creation of new jobs within the same firm for the operation and maintenance of the technology. A firm may also start supplying goods to new customers if input-saving technological change made the outputs more affordable, or if technological change affected the quality of the output which expanded its range of applications. For instance, the introduction of computers at the workplace creates new complementary tasks related to programming, hard- and software maintenance, and data management. 

Depending on the level of aggregation, the reinstatement effect also refers to jobs created upstream or downstream the supply chain, i.e. jobs associated with the production of $X$. For example, the suppliers of capital or intermediate inputs required to operate the new technology may increase their demand for labor if $X$ is used more intensively. Downstream industries may expand output output if upstream innovation reduces prices of intermediates. Hence, the reinstatement effect exists if $\partial L / \partial A > 0$ for any $A = A^L, A^X, A^Q$.

Here, we screen the empirical literature on whether or not it provides supporting evidence for the existence of this effect. Again, the measurement is complex as the technology-induced reinstatement of new jobs may happen at different levels of aggregation, i.e. in the same firm and/or in other industries. Hence, studies limited to a subset of firms or industries cannot capture the reinstatement of labor elsewhere. 

An increasing demand for labor is the key indicator of supporting evidence for the reinstatement effect. This is reflected in lower unemployment, an increasing number of employees, and hours worked. 
Note that the reinstatement effect does not need to be equally distributed across different types of labor and may co-exist with the replacement effect. To support the existence of this effect, it is sufficient if we observe an increase for at least one group. We also consider changes in the relative demand for labor as suggestive evidence for the existence of the reinstatement effect, as it may be driven by an increase in the demand for certain types of labor. 

Whether the net impact of technology on employment is positive or negative depends on the balance between labor replacement and reinstatement.

\subsection{The real income effect}
The two effects introduced above mainly refer to the direct impact of technology on the production side when it changes the use of inputs in absolute and relative terms. Technological change also affects labor through an indirect channel that mostly operates through the demand side. 

Assuming rational technology adoption decisions, technological change is always associated with productivity improvements; otherwise it would be irrational to adopt a new technology. Productivity improvements enable firms to produce a given value of output at lower costs which would be reflected in lower consumer prices $P$ if input costs savings are transmitted to consumers. Moreover, if technological change raises the marginal product of certain types of labor, we expect wages $w$ to rise. If technological change raises the marginal product of capital $K$, we also expect higher rents to capital which are another source of income. All these effects (lower prices $P$, higher wages $w$, higher returns to capital $r$) contribute to a rise in real income $I = \tfrac{wL + rK}{P}$. If demand is elastic and positively responds to increases in income ($\tfrac{\partial Q}{\partial (wL+rK)} \geq 0$) and decreases in prices ($\tfrac{\partial Q}{\partial p} \leq 0$), we can expect to observe an expansion of aggregate output $dQ \geq 0$. 

However, it should be noted that the real income gains are not necessarily equally distributed. This may have an impact on the demand reaction as the propensity to consume is heterogeneous across income groups and products. 
The expansion of output driven by a technology-induced real income effect may lead to a higher demand for labor. 

As the real income effect on labor is very indirect, we interpret a study as empirically supporting the real income effect if it provides empirical evidence for at least one of the underlying mechanisms, namely: an increase in (1) productivity, (2) lower prices, (3) higher levels of income and wages, and (4) rising levels of output and a positive relationship between labor and output. 

We interpret studies that report insights on at least one of these mechanisms as supportive for the real income effect while being aware that support for one of these mechanisms does not necessarily imply that the full chain of arguments holds. For example, productivity gains may not be forwarded to consumers in terms of lower prices if distorted competition prevents this, and rising levels of income do not necessarily imply a higher demand for consumption.

\section{Methods}
\label{sec:methods}
The aim of this review is to answer the question: \emph{What is the net employment effect of technological change since the 1980s in developed countries?} To answer this question, we reviewed the empirical evidence of studies published between 1988 - April 2021. 

\subsection{Search strategy}
We closely followed the PRISMA 2020 guidelines to ensure the quality of the systematic search process \citep{page2021litreview}. A scoping review was used to identify relevant search terms in widely cited studies. 
Subsequently, a computerized search was performed using the search terms that appeared either in the title, abstract, or list of keywords of studies, namely, i.e. `technolog*' combined with `labo\$r' or `employment'. The search was conducted in the Web of Science (WoS) Core Collection database.\footnote{A detailed description of the strategy and exact search strings used can be found in Appendix~\ref{appendix:searchstrategy}.} 


The initial search resulted in 8,748 studies published between 1 January 1988 and 21 April 2021. 
Given the large search outcome, six independent researchers were initially involved in screening the relevant records to ensure the timely and unbiased completion of the process. Based on the title and abstract,  studies were considered relevant if the independent variable is related to technology and the dependent variable is related to (un)employment. 
As a next step, the remaining 252 studies were assigned to three researchers, the authors of this paper, who independently assessed an equal amount of the remaining records based on the following inclusion criteria: (1) the study examines the effect of technological change on employment; 
(2) the study makes a significant empirical contribution, i.e. excluding purely theoretical studies; and (3) the study covers at least one developed country and the period after the 1980s.

Overall, the systematic search led to the inclusion of 127 studies. These studies were coded along a scheme that was iteratively refined throughout the process. We recorded the countries studied, the period covered, the outcome variable(s), and extracted descriptive information about the empirical operationalization of technological change, the measurement of the employment effects, the insights that can be gained about the three effects (replacement, reinstatement, and real income), and, if applicable, information about the net employment effect of technologies. We also extracted information about the level of analysis (e.g. country, region, firm, employee, occupation), data sources, sample size, methodology, potential heterogeneity, if applicable, and bibliometric information about the author, publication year and outlet.

\subsection{Coding scheme}
Throughout the process of screening the selected studies, we developed a simplified coding scheme for the effect considered, technology type, method, and level of analysis to cluster studies relying on similar measures and approaches. The coding scheme consists of categories (e.g. different technology types) and is based on common thematic patterns across the papers.

First, we classified a study as ``support''  if it provides results that offer data-based support for the existence of the effect of interest examined, i.e. replacement, reinstatement, and real income. Some studies find that the effects vary depending on the type of technology or the (sub)sample analyzed, for example distinguishing between industries, demographic groups, occupations, and types of labor. In this case we assigned the study to the category ``depends''. Studies that report negligible effects in terms of the magnitude or effects of low statistical significance were classified as ``weak''. Finally, studies were labeled as ``no support'' if they find that the effect of interest is insignificant or opposite to what was hypothesized.
Note that a study can investigate more than one type of effect, but the assigned labels, i.e. support, depends, weak or no support, are mutually exclusive within each effect. In the same spirit, we classify studies as ``positive'', ``negative'', ``depends'', and ``weak'' based on the extent to which they provide evidence on net employment effects.

The studies in our sample differ by the technologies examined and the empirical indicators used to measure them. We identified five broader classes of measures of technological change: ICT, robots, innovation, TFP-style, and a residual category other/indirect. The residual category accounts for technology types that are used by a small number of studies and are rather too heterogeneous to form separate technology groups.  

The increased availability of ICT was one major technology trend of the post-1980s. 
Empirical studies investigating its economic impact rely on different measures of ICT-diffusion such as 
ICT investments or capital, as for example found in \citet{EUKLEMS2019data} and other comparable databases which are mostly publicly available. Other studies rely on survey data of computer use and ICT investment at the firm-level or occupation-level. 

Another technology that raised much attention in economic research is the impact of robots. 
In most studies, robot-diffusion is measured using data for industrial robots from the IFR \citep{IFR2020data} which, to our knowledge, is the only source consistently covering a large set of countries, time-periods and all relevant industries where industrial robots are adopted. A few other studies use country-specific data from the Japanese and Danish national associations for robot manufacturers on robot sales \citep{dekle2020robot, graetz2018robots}, or trade in robots \citep{blanas2019who}. Two studies rely on survey data about industry level robot use \citep{camina2020autom, edler1994tech}, and one study relies on survey data on self-reported job losses in response to the adoption of robots at the workplace \citep{dodel2020perceptions}. Robots received much attention because they can be interpreted as a pure automation technology, substituting human workers performing manual tasks.

Next to these direct technology-diffusion measures, we group studies relying on innovation as indicator of technological change. These studies often use data from the Community Innovation Survey (CIS) or comparable surveys for non-European countries. These surveys are regularly conducted by statistical offices to assess the innovativeness of firms and regions. Typically, the surveys allow to distinguish between process and product innovation and, in some cases, organizational innovation. 
Process innovation is measured by survey questions asking firms to report whether they implemented a new improved production method, and product innovation is evaluated based on a question asking firms whether they recently introduced a new product. 
Organizational innovation is measured through a question asking for the implementation of new organizational methods in business practices, workplace organization or external relations \citep{arundel2013history}.

As a fourth technology cluster, we code studies as TFP-style if they rely on measures of technological change that are inferred from production functions and input uses. These measures are for example estimates of productivity (TFP or labor productivity) or changes in substitution elasticities. 
Some studies use readily available estimates provided by statistical offices or other relevant external data sources. Others explicitly estimate productivity or substitution elasticities on the basis of otherwise unexplained variation in the production function and sometimes also distinguish between different forms of input biases of technological change. 

As a remainder, we use the residual category other/indirect to label studies. This category comprises measures that are only used by a small number of papers. Some of these studies rely on indirect indicators measuring the risk of automation and exposure to technological change in certain industries, demographic groups, and regions. These studies often rely on metrics developed by \citet{frey2017future} which provide estimates of the probability that tasks in certain occupations performed by human labor can be automated. 
Another set of papers that we code as Other/indirect use trends in capital, high-tech equipment, and R\&D investments, and patents as proxies for technological change.  
This category also includes a variety of other measures such as shifts in the age of the capital stock, assuming that more recent capital indicates more dynamic patterns of technological change, or changes in the occupational efficiency. 

To cluster the papers by level of aggregation, we introduced four levels of analysis: Macro, meso, micro, and regional. Macro-level studies rely on country-level data and variation over time and/or countries. Meso-level studies usually include industry or sector-level data, while micro-level studies are at the more granular level ---mostly at the firm or individual employee-level. Finally, Regional level studies use variation across regional dimensions (e.g. commuting zones, NUTS regions, counties, etc.).

To classify papers by their methodology, we used three categories distinguishing between descriptive and conceptual analyses that link macro-level stylized facts to empirically reported technology trends (labeled ``descriptive''); regressions and similar forms of inferential statistics (labeled ``regression''); and studies that rely on simulation or calibration exercises (labeled ``simulation''). Simulation studies were only included if a substantial part of the study includes a significant amount of empirical data analysis, e.g. to motivate, calibrate, and/or estimate a simulation model. 
Some papers are classified into multiple categories, e.g. when more than one type of technology is studied, or when the authors performed the analyses on different levels of aggregation.

\begin{table}[H]
	\centering
    \caption{Studies across various categories}
    \label{tab:categories}
	\begin{tabu}{p{0.3\textwidth}*4{c}} \toprule
	\csname @@input\endcsname{inputs/stats_byeffect_short}
	\end{tabu}
	\begin{tabu}{l*{6}{c}} \toprule
	\csname @@input\endcsname{inputs/stats_bytech_short}
	\bottomrule
	\end{tabu}
	\begin{tabu}{p{0.35\textwidth}*{5}{>{\centering}p{0.095\textwidth}}} 
	\csname @@input\endcsname{inputs/stats_bylevel_short}
	\bottomrule
	\end{tabu}
    \begin{tabu}{p{0.3\textwidth}*{3}{>{\centering}p{0.19\textwidth}}} 
	\csname @@input\endcsname{inputs/stats_effect_bymethod_short}
	\bottomrule
	\end{tabu}  
	
	\begin{tablenotes}[flushleft] 
	\scriptsize \item Notes: Panel A presents the share and number (\#) of studies examining the replacement, real income, and reinstatement effect, respectively. The last column presents the share and \# of studies exploring at least two of the effects in the previous columns.
	Panel B presents the share and number (\#) of studies examining ICT, robots, innovation, TFP-style and Other technology groups, respectively. `Other' refers to technologies measured indirectly through prices, automation risks, etc. The last column presents the share and \# of studies examining at least two of the technology groups in the previous columns. Panel C presents the share and number (\#) of studies where the analysis is conducted at the macro (e.g. country, over-time), meso (e.g. sectors, industries), micro (e.g. firms, individuals), and regional (e.g. regions, states, cities) level of data aggregation, respectively. The last column presents the share and \# of studies where the analysis is conducted in at least two of the previous levels of analysis. Panel D presents the share and number (\#) of studies by the primary type of empirical methodology used in each study to identify the effect(s) of interest. `Descriptive' refers to studies using descriptive statistics and conceptual analyses that link macro-level stylized facts to empirically reported technology-trends at the micro-level. `Regression' refers to any regression-based analysis or other quantitative inferential methods with empirical foundation. `Simulation' captures simulation methods, e.g. DSGE. Note that there is no overlap in methods reported, i.e. more than one primary method used in each study, and thus the shares across columns add up to one, up to rounding. The total \# of studies is 127. 
	\end{tablenotes}
\end{table}

\section{Results}
\label{sec:results}
We begin with an overview of the basic descriptive statistics of the full sample of 127 studies. Subsequently, we describe and contextualize three subsets of studies that report empirical results for each of the three effects: Replacement, reinstatement, and real income (see Section \ref{sec:conceptual_framework}). 

\subsection{Overview}
Panel A in Table \ref{tab:categories} provides an overview of the studies by effect covered. The vast majority of studies (81\%) is related to the replacement effect. Another 62\% report results about the reinstatement and 26\% about the real income effect. More than half of the studies (59\%) simultaneously report results for at least two of the three effects and only 16 studies (13\%) for all three effects.

Panel B in Table \ref{tab:categories}, shows that roughly one third (35\%) of the studies analyzed the impact of ICT on employment. 13\% Studied the impact of robots and another 13\% examined the impact of innovation. About 14\% rely on TFP-style measures and 30\% fall into the residual category other/indirect. The distribution of Macro, Meso and micro-level studies is roughly balanced with 35\%, 32\%, and 30\%, respectively, while only 12\% have a regional focus. More than 80\% rely on regressions, 14\% on descriptives, and only 6\% used simulations. This result holds irrespective of the effect explored (see Appendix Table~\ref{tab:effect_bymethod}).

\begin{figure}[H] 
\caption{Share of studies by type of result reported for each effect examined}
\label{fig:share_results_reported}
\includegraphics[width=1\textwidth]{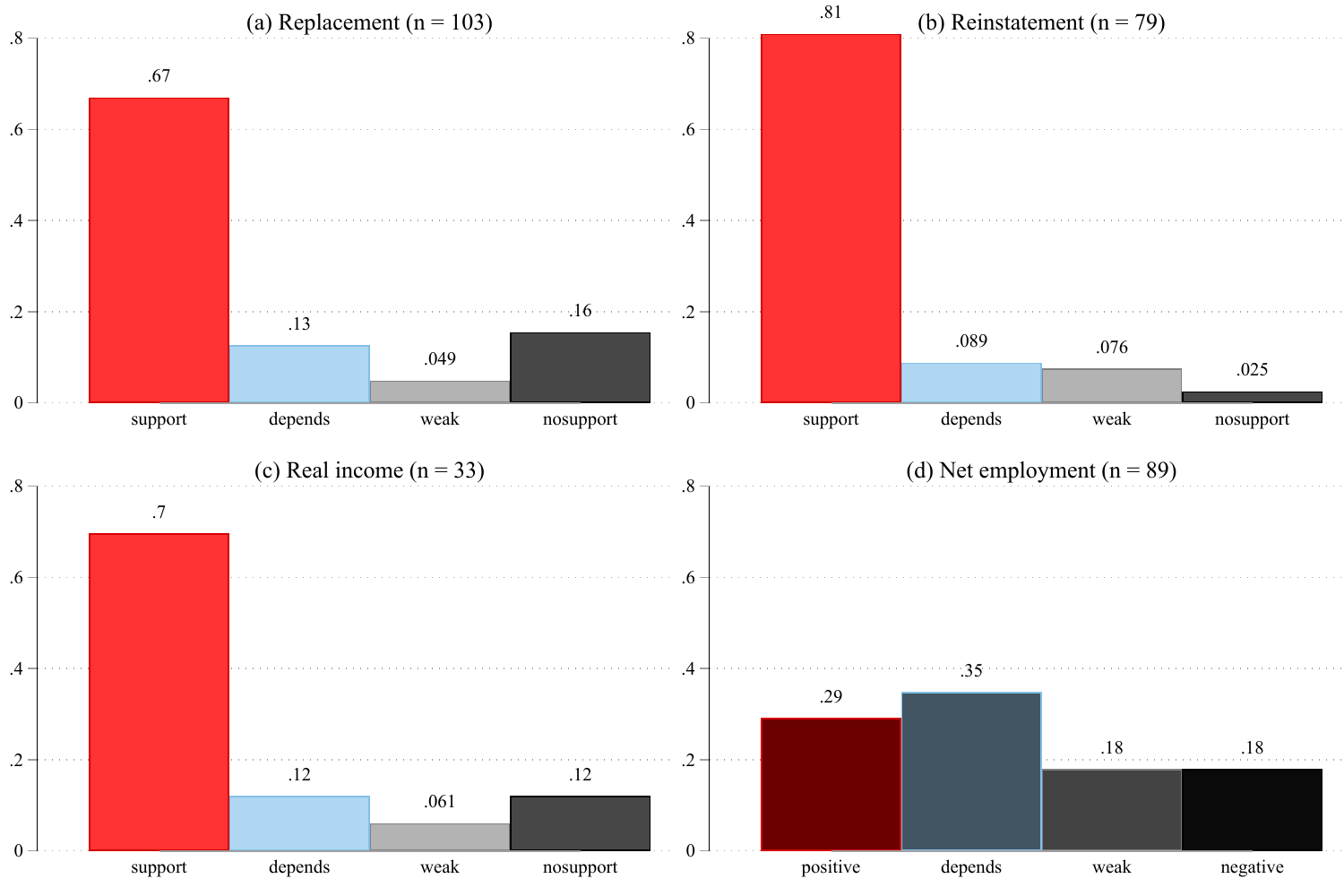}

\scriptsize 
Source: Author's calculations based on 127 studies collected from systematic literature review.\\
Notes: Panels (a)-(d) present the share of studies by each type of result reported for the replacement, reinstatement, real income and net employment effect, respectively. Specifically, in panels (a)-(c) a study is classified as `support' if it finds a significant empirical effect that supports the effect of interest examined, i.e. replacement, reinstatement, real income, respectively. `Depends' refers to the set of studies that find varying effects depending on the type of technology or the (sub)sample analyzed (e.g. type of sector or labor). Studies reporting effects that are negligible in terms of the magnitude were classified as `weak'. Studies were labeled as `no support' if they investigate the effect of interest, but documented insignificant or opposite effects. Similarly, in panel (d), `positive' and `negative' refer to studies which find for net employment a significant empirical effect that is positive and negative, respectively, while `depends' and `weak' are defined similarly to those above. A study can investigate more than one type of effect, but the assigned groups of results (i.e. support, depends, weak or no support and positive, depends, weak or negative) are mutually exclusive within each effect explored. In each panel, `n' is the number of studies examining the relevant effect. 
\end{figure}

In Figure \ref{fig:share_results_reported}, we provide an overview of whether or not the empirical findings of the studies support an effect. Roughly two-thirds (67\%) of the studies that report results on the replacement effect find support for this effect, while only 16\% provide no support. 

Panel (b) shows that among those studies that report results on the reinstatement effect, 81\% support the idea that technological change also creates new jobs either within the same firm, the same industry or elsewhere in the economy. Only a small fraction of the studies (2.5\%) find no support. 

The share of supporting studies (70\%) is similarly high for the real income effect (Panel (c)). It should be highlighted that the number (n) of studies in this subsample is much smaller compared to the rest of the effects, i.e. 33 versus 103 and 79 studies.\footnote{Note that the real income effect interacts with some studies that rely on positive productivity shocks (TFP-style). These studies assume that productivity increases are an indicator of technological change. Our search for support of the real income effect asks whether this assumption holds true and we aim to understand the channels through which productivity interacts with demand as potential source of labor reinstatement.} 

Panel (d) in Figure \ref{fig:share_results_reported} summarizes the results from papers that provide empirical evidence on the net effect. In total, 29\% of studies document a net positive effect, 18\% a negative effect, and 18\% report ambiguous or inconclusive results. The relative majority of studies (35\%) shows that the net employment effect depends. 

\FloatBarrier
\subsection{Replacement effect}
\label{sec:results_replacement}

\subsubsection{Overview, methods and technical issues}

Does technology replace human labor? The majority of studies exploring the Replacement effect suggests that it does, but we also find a relatively small share of studies that do not support or suggest ambiguous effects (see Table \ref{tab:replacement}). Here, we systematize the empirical evidence by result (support, no support, depends, weak) and other characteristics of each study. 

\begin{table}[H] \centering \footnotesize 
\begin{threeparttable}
\caption{Studies by findings on replacement}
\label{tab:replacement}
    \begin{tabular}{l*{4}{c}} \toprule
    & (1) & (2) & (3) & (4) \\
	\csname @@input\endcsname{inputs/stats_replace_byfind}
	\bottomrule
	\end{tabular}
	\begin{tablenotes}[flushleft] 
	\scriptsize \item Notes: Columns (1)-(4) present the share and number (\#) of studies with empirical results that support, depend (on various characteristics, e.g. technology type, analysis level, etc.), are weak, and do not support the presence of a  replacement effect, respectively. The total \# of studies in the sample is 127 of which 103 examine the replacement effect.
	\end{tablenotes}
\end{threeparttable}	
\end{table}

Many studies build on the neoclassical framework of skill- or task-biased technological change as a basis for the empirical analysis. 
The majority of studies (83\%) use regressions, fifteen studies (15\%) rely on descriptive analyses, and the remaining three studies (3\%) use simulation methods (see Appendix Table \ref{tab:effect_bymethod}). 

In column (1) of Table \ref{tab:tech_byfinding_replacement}, we present the fraction of studies by technology type (ICT, robots, innovation, TFP-style, other/indirect). The other columns display for each technology whether or not empirical support for the replacement effect is found. 
Most studies (36\%) examine the impact of ICT, followed by TFP-style (17\%), robots (15\%), and innovation (12\%). 29\% use other measures of technology that fall in our residual category other/indirect. 

\begin{table}[H] \centering \footnotesize 
\begin{threeparttable}
\caption{Studies by findings on replacement effect for each technology group considered}
\label{tab:tech_byfinding_replacement}
    \begin{tabular}{lc|*{4}{c}} \toprule
    & (1) & (2) & (3) & (4) & (5) \\
    & \multicolumn{1}{c}{}& \multicolumn{4}{c}{by finding} \\
    \cmidrule{3-6}	
	& \multicolumn{1}{c}{total}& \multicolumn{1}{c}{support}&\multicolumn{1}{c}{depends}&\multicolumn{1}{c}{weak}&\multicolumn{1}{c}{no support}\\ \midrule
	\csname @@input\endcsname{inputs/stats_tech_byfind_replace}
	\bottomrule
	\end{tabular}
	\begin{tablenotes}[flushleft] 
	\scriptsize \item Notes: 
	Column (1) presents the share and number (\#) of studies by the technology group considered in each row panel relative to the total number of studies examining the replacement effect. The row-sum of shares in column (1) does not add up to one since there are studies considering more than one technology group. Columns (2)-(4) present the share and \# of studies by the set of findings reported on the replacement effect for each type of technology considered in each row panel. For findings, a study is classified as `support' if it finds a significant empirical effect that supports the replacement effect examined. `Depends' refers to the set of studies that find varying effects depending on the type of technology or the (sub)sample analyzed (e.g. type of sector or labor). Studies reporting effects that are negligible in terms of the magnitude were classified as `weak'. Studies were labeled as `no support' if they investigate the effect of interest, but documented insignificant or opposite effects. The technology types reported include ICT, robots, innovation, TFP-style and Other types of technologies (e.g. indirectly measured through prices, automation risks), respectively.  Note that there is no overlap in findings reported, i.e. more than one primary finding in each study, and thus the shares across columns add up to one, up to rounding. The total \# of studies in the sample is 127 of which 103 examine the replacement effect. 
	\end{tablenotes}
\end{threeparttable}	
\end{table}

Column (2) in Table \ref{tab:tech_byfinding_replacement} shows high support rates across all technologies except for innovation. The strongest support comes from studies that use robots (87\%), TFP-style (76\%), other/indirect (73\%), and ICT (62\%). 
The findings from ICT studies are the most controversial showing simultaneously high numbers of supporting (62\%) and non-supporting studies (24\%), but only a few studies that report conditional or weak effects. 
Innovation studies show the weakest support rate for the replacement effect, but seem consensual about the conditionality of the effect. Only one-quarter (25\%) supports this effect while 17\% does not find any empirical support. The majority (58\%) reports that the effect is conditional on the type of innovation (e.g. product or process innovation) and other relevant dimensions, such as the characteristics of the employees and firms.

\FloatBarrier
\subsubsection{Studies supporting the replacement effect}

The highest absolute number of papers (n=23) supporting the replacement effect studies the impact of ICT \citep{autor2002computers,autor2003skill,autor2013polar, baddeley2008structural,balsmeier2019time,dengler2018tech, diaz2002inno, eden2018, fonseca2018polar, fossen2021digital, morrisonpaul2001tech, wolff2009computer, addison2000tech, autor2015there, berman1994changes, luker1997employment, autor2015trade, blanas2019who, goaied2019, kristal2013capitalist, kaiser2001, jerbashian2019autom, morrison1997}. 
These papers mostly build on theories of skill- or task-biased technological change and suggest that ICT replaces human labor in low-skill jobs and occupations or (regions with) industries that are intensive in routine tasks. Most studies are at the meso-level of analysis, followed by studies at the micro-level \citep{balsmeier2019time, dengler2018tech, fonseca2018polar, fossen2021digital, addison2000tech, kaiser2001} and macro-level \citep{baddeley2008structural,eden2018,wolff2009computer,autor2015there,goaied2019}. Only \citet{autor2013polar, autor2015trade} found ICT-induced replacement effects at the regional level. 

The highest relative support (87\%) for the replacement effect comes from studies on robots (n=13) \citep{borjas2019, camina2020autom, devries2020rise, edler1994tech, faber2020robot, jung2020tech, compagnucci2019robotization, acemoglu2019automation, acemoglu2020robots, dodel2020perceptions, blanas2019who, graetz2018robots, labaj2020automation}. 
Amongst these, \citet{compagnucci2019robotization} reported a positive effect on wages which is outweighed by employment losses, yielding a net negative effect on the wage bill. \citet{faber2020robot, acemoglu2020robots} studied the impact of robots on the regional (commuting zone) employment-population ratio. 
\citet{acemoglu2020robots} used a long difference regression approach and report a negative relationship between robot exposure and employment where the effects were strongest in routine manual occupations and blue collar jobs. \citet{borjas2019} rely on similar data-sets to examine the impact of robots and immigrants on hourly wages and employment at the state-industry-level. They observed a negative association between robots and employment and wages. 

\citet{faber2020robot} examined the impact of robot usage in an offshoring country. He found that robot adoption in offshoring industries in the US had a negative employment effect on regions in Mexico that heavily rely on exports to the US. This result supports the idea of cross-regional spillovers. When regions' comparative advantage is based on low labor costs in the production of tradable goods, they might lose this advantage as the tasks performed by cheap labor can now be performed by machinery. 

One macro-level robot study examined the impact of robots, measured by the number of robots per thousand workers, on changes in the country-level wage bill \citep{labaj2020automation}. The authors found evidence for the existence of the replacement effect, but simultaneously reported that negative employment effects are overcompensated by the reinstatement of new labor.

With only three studies, the empirical support for the replacement effect based on innovation studies is weak \citep{vivarelli1996inno,cirillo2017tech,dachs2017inno}.
\citet{vivarelli1996inno} studied the impact of product and process innovation in Italian manufacturing firms during the 1980-1990s. They found support for labor displacement driven by the dominant role of process innovation. 
\citet{cirillo2017tech} and \citet{dachs2017inno} made the same observation in different European countries, but also highlighted the presence of heterogeneity in their findings with stronger effects in high-tech industries and in Northern Europe.

Thirteen TFP-style studies support the Replacement effect empirically, mostly reporting shifts in labor demand, particularly across industries and across different types of labor e.g. from production to non-production labor  \citep{angelini2020wage, ergul2020tech, gregory2001jobs, ho2008invest, baltagi2005skill, bloch2010technological, chen2014total, autor2018lshare, bessen2020autom, graham2000, whelan2009shocks, huh2019, kim2020lshare}. 
\citet{angelini2020wage, ho2008invest, baltagi2005skill, gregory2001jobs} found that technological change is skill-biased by showing that lower skilled labor tends to be replaced by high-skilled labor. 
\citet{bessen2020autom, chen2014total, bloch2010technological} evaluated the factor bias of technological change and documented patterns of labor-saving and capital using technologies over the past decades with considerable heterogeneity across industries and countries. \citet{autor2018lshare} documented that industry-level TFP-growth is associated with a decrease in wages, hours worked, the wage bill, and the labor share in major industrial countries since the 1970s. 
\citet{whelan2009shocks, huh2019} studied cyclical macroeconomic fluctuations showing positive productivity shocks to be negatively related to hours worked. \citet{ergul2020tech, kim2020lshare} suggested that technology-induced shocks are associated with decreases in the labor share of income, albeit potentially being only transitory. \citet{graham2000} found that some of the industry-region employment losses can be attributed to technological change.

The studies using other/indirect technology measures (n=22) can be roughly grouped into three categories. A first bundle of studies \citep{peng2018it, gardberg2020digit, blien2021occup, grigoli2020automation, arntz2017} exploited variation in regional, industrial and/or occupational susceptibility to automation and provide evidence that automation risk indices help to explain employment losses and the longer unemployment spells for workers in certain jobs. 

A second category of other/indirect technology measures includes capital and high-tech equipment investments as proxies for technological change \citep{morrison1997, wemy2019lshare, ho2008invest}. Also \citet{gera2001tech} infer the pace of technological change from shifts in the age of the capital stock, as more recent capital indicates more dynamic patterns of technology investment. 
\citet{kim2020lshare, gera2001tech, morrisonpaul2001tech, vainiomaki1999tech} approximated technological change via R\&D spending and \citet{gera2001tech, feldmann2013} used patents. 
Furthermore, \citet{gera2001tech} provide support for skill-biased technological change, i.e. more negative consequences in terms of employment and income for low-skill labor. 
The skill bias is also supported by the descriptive analyses of \citet{oesch2010, buyst2018polar} made a detailed description of occupational shifts across industries attributing them to technology trends. Similarly, the detailed historical analysis from \citet{padalino1997employment} documents changes in the employment-GDP elasticity in G7 countries for the period 1964-94 and found - for the late 20th century - a negative correlation between GDP growth and employment in manufacturing, but not for the whole economy. 

The third category of studies focuses on skill-biased technological change. They are all at the macro-level except for one regional level study \citep{reshef2013tech, padalino1997employment, manning2004tech, hoskins2000, hoskins2002, madariaga2018employment, reijnders2018tech}. The measures of technology employed are heterogeneous and vary from the use of the Leontief inverse matrix as a proxy for changes in production techniques to changes in occupational efficiency. All studies provide suggestive evidence that technological change decreases the (relative) demand for unskilled workers or routine jobs.


\subsubsection{Studies not supporting the replacement effect}
The number of studies that provide evidence against any significant impact of technology on employment is somewhat smaller (n=17). Nine of them are on ICT, two on robots, two on innovation, one on TFP-style study, and three studies use other/indirect technology measures.  

\citet{fung2006} proxied ICT by expenses on IT and computer data processing and examined whether ICT had a labor-saving impact on firms in the US banking industry between 1992-2002. The author rejected the presence of labor-saving effects for the technologies considered, observing that the more technology-intensive firms increased their employment. Similarly, \citet{auberttarby2018} analyzed whether digitilization in the French newspaper and magazine press industry creates or destroys jobs. 
Overall, they found digitalization to be associated with higher wages and a reduced probability to be laid off. 

\citet{pantea2017ict, atasoy2016, gaggl2017computer} made similar observations for EU countries and Turkey looking at different firm-level ICT usage indicators. They estimated either insignificant or positive effects of these technologies on employment in SMEs. \citet{gaggl2017computer} showed that the introduction of these technologies is associated with changes in the composition of tasks performed by workers. 
In line with that, \citet{biagi2017ict, ivus2015broadband} did not find any significant effect of ICT on employment in European industries and Canadian regions, respectively. 
\citet{behaghel2016it} did not observe that ICT-diffusion increases the probability of dismissals in French cities. If anything, they provided only weak support of increased job instability for high-school dropouts.
\citet{borland2017robot} provided a descriptive historical analysis documenting a decrease in routine labor which does not deviate from historical patterns. They could not identify any noteworthy relationship between ICT-diffusion and employment changes.

\citet{dekle2020robot} and \citet{fu2021} evaluated the impact of robots on employment (n=2). 
\citet{fu2021} studied a sample of seventy-four countries, including the EU-27 and twenty-nine developing countries. They found insignificant effects in developing countries and positive employment effects accompanied with productivity increases in developed countries. The authors also reported heterogeneity by gender, but they did not find evidence for robot-induced replacement. \citet{dekle2020robot} analyzed the impact of robot-diffusion during 1979-2012 in twenty Japanese industries, but could not identify any significant negative effects on employment. 

Among innovation studies (n=2), \citet{calvino2018innovation} concluded in a literature review that the relationship between product innovation and employment is rather positive if statistically significant at all. The impact of process innovation remains controversial and the authors emphasized that the existing evidence is insufficient to support the existence of a replacement effect. \citet{fung2006} studied - in addition to ICT - the impact of in-house process innovation in the banking sector. But again, they could not find any support that these innovations reduce employment. 

Three studies considered other/indirect technology measures. Using regressions, \citet{scholl2020autom} did not find any significant impact of automation risks on pay or employment. 
\citet{sargent2000unempl} assumed that changes in economic regularities during the 20th century, such as the relationship between unemployment rates and vacancies, are an indication of technological change.
The author found descriptive evidence that such shifts occurred and linked these patterns to employment shifts across industries, occupations and educational groups, but he could not find any evidence of increased unemployment.
\citet{raval2019micro} explored plant-level information to examine whether capital-labor substitution possibilities have changed between 1987-2007. They found that capital-labor elasticities have been very persistent over the time period covered.

\subsubsection{Studies with ambiguous findings and indirect evidence}
Ambiguous results are reported in twenty studies coming in the following decreasing order from innovation (n=7), ICT (n=5), other/indirect (n=5), and TFP-style (n=3). 
No robot study reports ambiguous results. 
Among the studies with ambiguous results, some of these---classified as ``weak''---provide only indirect evidence that may be consistent with the replacement effect. The remaining studies---classified as ``depends''---provide contradicting evidence when using different specifications. In both cases and irrespective of the technology, all of them remain inconclusive on whether or not technology replaces labor. 

%

Among the ICT studies, \citet{gallipoli2018it} provided indirect support for the replacement of non-IT by IT-intensive occupations with a higher wage premium. 
\citet{breemersch2019polar} showed ICT adoption to be associated with increased polarization towards high pay jobs within manufacturing only, but with minimal contributions between manufacturing and non-manufacturing industries. 
\citet{maurin2004skill} found that the share of skilled labor increased in response to ICT adoption which may be accompanied with the simultaneous replacement of unskilled labor. 
\citet{autor1998computers} observed rapid skill upgrading processes since 1970 and provided suggestive evidence that these processes evolve faster in computer intense industries. 
These studies point to a declining employment share of certain types of jobs, i.e. mostly routine task intensive, low-skill, and low-wage jobs. However, it is not necessarily true that changes in employment shares are accompanied by the replacement or ``phasing out'' of certain occupations. 
\citet{omahony2021lshare} studied the impact of ICT and innovation-related capital investments, but did not find any conclusive results on whether the labor share is positively or negatively affected. Intangible investments related to innovation tend to have a positive impact, while investments related to firm organization tend to show the opposite pattern. Low- and middle-skill workers appear to be more negatively affected. 

Three studies of those relying on TFP-style measures report ambiguous results \citep{boyle2002trade, dixon2020decline, fort2018}. 
\citet{fort2018} made a detailed descriptive analysis showing a negative relationship between TFP-growth and employment in some but not all manufacturing industries. They reported that this does not realize through the shut-down of existing plants but through lower shares of labor input in plants that are new entrants. 
\citet{boyle2002trade, dixon2020decline} did not directly evaluate the impact of TFP on labor, but showed that the decline in the labor share can be partly attributed to labor-saving technological change. 

Seven studies analyzed the impact of innovation \citep{bogliacino2010inno, falk2015inno, cozzarin2016inno, evangelista2012inno, pellegrino2019rnd, vanreenen1997inno, kwon2015inno}. 
All of them are firm-level analyses - except for the industry-level study by \citet{bogliacino2010inno} - and explored whether there are any effects of process/product innovation on labor demand. \citet{falk2015inno, evangelista2012inno, kwon2015inno} also studied the impact of organizational innovation. None of these studies could identify any significant impact of product and organizational innovation on labor. For process innovation, \citet{pellegrino2019rnd, vanreenen1997inno} either found only weakly significant negative effects or showed that the result depends on the size of firms. 
\citet{falk2015inno} showed a negative association between process innovation and labor, but only for a subset of industries. 
\citet{cozzarin2016inno} found ambiguous results depending on the type of innovation and technology considered: For example, he found no effect of advanced manufacturing technologies, a negative effect of product innovation, and a positive effect of process innovation. 
\citet{evangelista2012inno} found weak evidence of a labor-saving impact of organizational product innovation in manufacturing but not in services. Next to the labor-saving effect, they also observed an innovation-induced increase in sales which may offset the negative effect on employment. 
\citet{kwon2015inno} reported a negative impact of process but not product innovation. 

Among those studies using other/indirect technology measures (n=5), three used investment in high-tech capital or R\&D as technology proxies \citep{breemersch2019polar, flug2000, idris2021tech} and \citet{omahony2021lshare} used knowledge-based assets. \citet{idris2021tech} could not identify any impact of high-tech on employment and \citet{flug2000, breemersch2019polar} documented technology-induced wage and skill polarization patterns in the labor market. However, \citet{breemersch2019polar} emphasized that technology plays a rather minor role in explaining labor market polarization. 
\citet{omahony2021lshare} found heterogeneous results depending on the type of the technology considered: R\&D-based knowledge investments seem to mitigate the ICT-driven declining trend in the labor share, while innovation-related intangible investments are related to a rising labor share, opposite to the effect found for organization-related investments.

Finally, \citet{green2015} provide a descriptive analysis of technology trends and explored polarization in the Canadian labor market since the 1970s. They cannot confirm any hypothesis about skill-biased technological change. Rather, the drivers of the polarization differ from those in the US and other countries. The authors documented some occupations to be shrinking but argued this to be driven by the resource boom in Canada, rather than by technological change. 

\FloatBarrier
\subsection{Reinstatement effect}
\label{sec:results_reinstatement}

\subsubsection{Overview, methods and technical issues}
Does the introduction of new technologies create new jobs? 
In total, seventy-nine studies in our sample offer empirical insights on the reinstatement effect. 
Among those studies that report empirical evidence on the reinstatement effect, 81\% support the existence of this effect, while 17\% report ambiguous findings and only two studies (3\%) document insignificant effects (see Table \ref{tab:reinstatement}). 
Almost all of the reinstatement studies (82\%) simultaneously report findings on the replacement effect and seventeen studies (22\%) on the real income effect.

\begin{table}[H] \centering \footnotesize 
\begin{threeparttable}
\caption{Studies by findings on reinstatement}
\label{tab:reinstatement}
    \begin{tabular}{l*{4}{c}} \toprule
    & (1) & (2) & (3) & (4) \\
	\csname @@input\endcsname{inputs/stats_reinst_byfind}
	\bottomrule
	\end{tabular}
	\begin{tablenotes}[flushleft] 
	\scriptsize \item Notes: Columns (1)-(4) present the share and number (\#) of studies with empirical results that support, depend (on various characteristics, e.g. technology type, analysis level, etc.), are weak, and do not support the presence of a  reinstatement effect, respectively. The total \# of studies in the sample is 127 of which 79 examine the reinstatement effect. 
	\end{tablenotes}
\end{threeparttable}	
\end{table}

In column (1) of Table \ref{tab:tech_byfinding_reinstatement}, we show that the largest fraction of studies looking at reinstatement focus on the impact of ICT (38\%) and other/indirect technology measures (28\%), followed by Innovation (18\%),  TFP-style (14\%), and Robots (13\%). 
Similarly to the replacement effect, the highest ambiguity (weak and depends) is observed for innovation (43\%) and ICT (20\%). However, even in these cases the support rates remain high. 

The vast majority of studies (n=63) relies on regressions, thirteen studies rely on descriptive analyses, and only three are based on simulations (see Appendix Table \ref{tab:effect_bymethod}). 

\begin{table}[H] \centering \footnotesize 
\begin{threeparttable}
\caption{Studies by findings on reinstatement effect for each technology group considered}
\label{tab:tech_byfinding_reinstatement}
    \begin{tabular}{lc|*{4}{c}} \toprule
    & (1) & (2) & (3) & (4) & (5) \\
    & \multicolumn{1}{c}{}& \multicolumn{4}{c}{by finding} \\
    \cmidrule{3-6}	
	& \multicolumn{1}{c}{total}& \multicolumn{1}{c}{support}&\multicolumn{1}{c}{depends}&\multicolumn{1}{c}{weak}&\multicolumn{1}{c}{no support}\\ \midrule
	\csname @@input\endcsname{inputs/stats_tech_byfind_reinst}
	\bottomrule
	\end{tabular}
	\begin{tablenotes}[flushleft] 
	\scriptsize \item
	Notes: Column (1) presents the share and number (\#) of studies by the technology group considered in each row panel relative to the total number of studies examining the reinstatement effect. The row-sum of shares in column (1) does not add up to one since there are studies considering more than one technology group. Columns (2)-(4) present the share and \# of studies by the set of findings reported on the reinstatement effect for each type of technology considered in each row panel. For findings, a study is classified as `support' if it finds a significant empirical effect that supports the reinstatement effect examined. `Depends' refers to the set of studies that find varying effects depending on the type of technology or the (sub)sample analyzed (e.g. type of sector or labor). Studies reporting effects that are negligible in terms of the magnitude were classified as `weak'. Studies were labeled as `no support' if they investigate the effect of interest, but documented insignificant or opposite effects. The technology types reported include ICT, robots, innovation, TFP-style and other types of technologies (e.g. indirectly measured through prices, automation risks), respectively.  Note that there is no overlap in findings reported, i.e. more than one primary finding in each study, and thus the shares across columns add up to one, up to rounding. The total \# of studies in the sample is 127 of which seventy-nine examine the reinstatement effect. 
	\end{tablenotes}
\end{threeparttable}	
\end{table}

\subsubsection{Studies supporting the reinstatement effect}

Among those studies that support the reinstatement effect, twenty-three look at the impact of ICT \citep{atasoy2013broadband,auberttarby2018,autor2003skill,autor2013polar,baddeley2008structural,balsmeier2019time,fossen2021digital,fung2006,morrisonpaul2001tech,behaghel2016it,autor2015there,berman1994changes,luker1997employment,autor2015trade,blanas2019who,gaggl2017computer,gallipoli2018it,ivus2015broadband,kristal2013capitalist,kaiser2001,jerbashian2019autom,maurin2004skill,morrison1997}. 

At the city-level, \citet{behaghel2016it} studied the impact of ICT on skill upgrading and job-to-job transitions showing that ICT adoption is associated with an increased demand for skilled labor, but not with higher dismissal rates. 
\citet{gaggl2017computer} and \citet{ivus2015broadband} argued that the impact of ICT on net employment depends on various dimensions. \citet{gaggl2017computer} found a positive relationship between ICT adoption and the demand for non-routine cognitive labor, but did not observe effects on job replacement. However, they found that the positive effect diminishes over time which may indicate that it is only a transitory phenomenon. 
The detailed descriptive analysis of industrial job creation and destruction dynamics by \citet{borland2017robot} provides evidence that the positive employment effects in Canada are heterogeneous across regions: ICT-diffusion only exhibits significant positive interactions with labor in rural regions with stronger effects in ICT-intensive industries. 

\citet{gallipoli2018it} listed numerous empirical stylized facts for the US between 1980-2013 using micro-level census data and detailed occupational employment statistics. They documented the emergence of new and well-paid IT occupations, mostly in services. They showed that productivity growth can be mostly attributed to services while the employment share in manufacturing declined. However, their findings do not allow to draw conclusions about the existence of simultaneous replacement or net employment effects.  
\citet{maurin2004skill} explored employee-level data on different types of computer technologies used at work. They found a positive correlation between the diffusion of computers and the share of high-skilled labor. 
The authors also showed that the impact of ICT on labor demand is conditional on the type of ICT, and on the occupations and their task content. 

Only the study by \citet{atasoy2013broadband} reports findings exclusively for the reinstatement effect, but not on any of the other two effects. The authors studied the impact of broadband deployment on county level labor markets in the US between 1999-2007. They found a positive effect of broadband access on employment and wages. The positive wage effects are larger in counties with more skilled labor supporting theories of skill-biased technological change. 

In an economic history essay, \citet{autor2015there} documented in detail how the demand for labor for certain occupations in services, such as managers, personal care, food services and others, continuously increased since the 1980s until the financial crisis when the patterns of growth slowed down. Furthermore, he illustrated patterns of skill polarization reflected in the highest growth rates in the lowest and highest skill percentile. 
Similarly, \citet{luker1997employment} documented shifts from manufacturing to services. They showed that a net increase in high-tech industries can be mostly attributed to services. Generally, high-tech service employment appears to grow faster than employment in the rest of the economy.  

Thirteen regression studies on ICT simultaneously support the replacement and reinstatement of labor \citep{autor2003skill,autor2013polar,baddeley2008structural,balsmeier2019time,fossen2021digital,morrisonpaul2001tech,berman1994changes,autor2015trade,blanas2019who,kristal2013capitalist,kaiser2001,jerbashian2019autom,morrison1997}. 
The two Regional level studies by \citet{autor2015trade, autor2013polar} found support for routine-biased technological change and observed increasing employment in abstract and manual tasks which neutralizes the negative employment effects of routine-task replacement.  
\citet{autor2013polar} also documented a rise in polarization, i.e. both a rise in low-skill service jobs and a differential wage growth across occupations. 
The study by \citet{baddeley2008structural} is a macro-level analysis confirming that computerization, next to financialization, was associated with employment shifts from manufacturing to services in the UK between 1979-2005. 

At the meso-level,  \citet{autor2003skill, morrisonpaul2001tech, berman1994changes,  blanas2019who, kristal2013capitalist, jerbashian2019autom, morrison1997} observed that ICT adoption is associated with the creation of new jobs. \citet{autor2003skill} found an increase in non-routine jobs. \citet{morrison1997} showed that investments in computers and R\&D are associated with skill polarization patterns, i.e. an increased demand for high- and low-skill workers. 
\citet{berman1994changes} documented a rise in non-production labor which coincides with a skill upgrading process. 
A similar skill bias is confirmed by \citet{blanas2019who} who showed that robots and ICT are associated with higher employment in high- and middle-skill jobs and in services. 
Relatedly, \citet{jerbashian2019autom} showed a positive correlation between falling IT prices and employment in high-wage occupations. 

 \citet{balsmeier2019time} and \citet{kaiser2001} provided evidence at the firm-level and \citet{fossen2021digital} at the level of individual employees.
The two firm-level studies found that increased investment in IT is associated with the creation of high-skill jobs. \citet{fossen2021digital} found evidence for the creation of another class of jobs showing that digitalization significantly increases the probability that high-skill workers engage in entrepreneurial activity. 

Nine reinstatement supporting studies analyzed the impact of Robots \citep{devries2020rise, dekle2020robot, edler1994tech, acemoglu2019automation, blanas2019who, graetz2018robots, leigh2019robots, gentili2020machines, labaj2020automation}. 
Two thirds of them reported ambiguous effects on net employment and the remaining three \citep{leigh2019robots, gentili2020machines, dekle2020robot} found positive effects. None of these studies documents a clear negative impact of robots on net employment. This suggests that whenever evidence for robot-driven reinstatement is found, it tends to (over)compensate the replacement of labor. 

\citet{edler1994tech} built on an empirical input-output model and found a higher demand for skilled labor in response to robot-diffusion. 
\citet{gentili2020machines} performed a descriptive clustering analysis and attributed changes in the robot intensity to changes in employment measured by hours worked finding that industries with the highest robot intensity experienced the highest productivity and employment gains. The effects are clustered in high-tech industries which account for a small share of total employment. 

\citet{labaj2020automation} offer a macro-level regression analysis exploring how variations in the economy wide wage bill and labor share are explained by robot-diffusion rates in the US and Europe. They decomposed aggregate changes in the task content of production into a reinstatement and replacement effect and found evidence for both. However, the authors emphasized that whether reinstatement or replacement dominates varies across countries (especially between the US and EU) and that this variation cannot be explained by robot-diffusion.

For six meso-level regression-based analyses \citep{devries2020rise, dekle2020robot, acemoglu2019automation, blanas2019who, graetz2018robots,leigh2019robots} we observed that they mostly report increases in high-skill and in service jobs. \citet{dekle2020robot} found similar effects for total employment, i.e. not only for high-skill and service jobs. \citet{leigh2019robots} also showed that in US manufacturing the impact of robots tends to be positive. 

Continuing, eight studies relying on innovation offer support for the reinstatement effect. 
\citet{calvino2018innovation} made a literature review on the impact of innovation, captured by different indicators (R\&D intensity, CIS, patents) on employment at the firm and sector-level. They documented that the employment effect is mostly positive, but also depends on the sector and the type of innovation (i.e. process or product). They confirmed a mostly positive effect of product innovation and a negative effect of process innovation. 
\citet{tether1998employment} descriptively explored employment creation by small innovative firms at the micro-level, where innovative firms are defined as inventor award winning firms. 
The authors showed that these firms have faster than average employment growth patterns, but it remains unclear to which extent this observation can be generalized beyond this specific setting. 

Three studies are micro-level regressions using firm-level data \citep{fung2006, vivarelli1996inno, dachs2017inno}. \citet{fung2006} studied labor-saving product and process innovation in the banking sector and found: (a) a positive association between labor-saving technologies and higher firm-level employment and (b) positive spillovers from patented process innovations on labor demand in non-innovating banks. The latter effect suggests the presence of technology-induced employment externalities to non-technology-adopting sectors and how these channels contribute to the creation of labor in the economy. 

Both \citet{vivarelli1996inno} and \citet{dachs2017inno} observed that product innovation is positively associated with sales and labor demand, but \citet{vivarelli1996inno} found this pattern only for a subset of sectors which are characterized by higher design and engineering intensities and higher percentages of product innovations. 

\citet{cirillo2017tech} and \citet{xiang2005} performed industry-level analyses. \citet{cirillo2017tech} found a positive relationship between the share of firms introducing product innovations, industry-level demand, and employment growth. 
Similarly, \citet{xiang2005} showed that the introduction of new goods is positively associated with the relative demand for skilled labor in the US manufacturing industry. 

\citet{capello2013inno} confirmed the positive association of product innovation with employment using regional-level data for the EU. Applying spatial regressions, they showed a positive relationship between the share of firms that introduce product innovations and regional employment and wage growth in sub-national regions (NUTS2) that are characterized by a high share of blue collar workers. The authors also highlighted that the effect in regions with low shares of blue collar workers may be negative, but with no insights on whether these relationships have changed over time. 

Ten studies rely on TFP-style measures. In a descriptive analysis, \citet{angelini2020wage} inferred technological change from shifts in the skill content of production and provided evidence for the reinstatement of service jobs reflected in employment shifts from manufacturing to services. \citet{fort2018} conducted a detailed descriptive analysis of the impact of labor productivity changes on employment shifts at the industry and plant-level emphasizing that displacement and net effects are industry-specific. 

The other eight papers are regressions, largely at the industry or macro-level. They rely on different proxies of productivity changes, mostly captured by TFP or labor productivity, and study labor demand in absolute and relative terms. \citet{bessen2020autom} investigated the role of the elasticity of demand with respect to productivity for three industries since the 19\textsuperscript{th} century. The author highlighted that final demand was historically a key driver of the reinstatement of labor in the steel, textiles and automobile industry. 
The studies by \citet{ho2008invest, baltagi2005skill, kim2020lshare, sala2018effects} showed that the reinstatement of labor may be biased, as reflected in an increasing demand for skilled and non-production labor. 
The other three studies provide similar findings in different settings. \citet{boyle2002trade} found that capital accumulation and technological change are key drivers of employment growth. \citet{autor2018lshare} investigated TFP shocks in upstream industries and showed that this is positively associated with hours worked and employment in downstream industries. \citet{graham2000} showed that regional employment increases can be attributed to technology. 

Twenty studies fall into the residual category other/indirect. 
Among these, \citet{morrisonpaul2001tech, feldmann2013, vainiomaki1999tech, vanroy2018tech, yildirim2020, kim2020lshare, fagerberg1997tech} used R\&D expenditures or patents as a measure of technological change. \citet{morrisonpaul2001tech} and \citet{kim2020lshare} observed a positive relationship between R\&D expenditures and employment of high-skill labor. \citet{yildirim2020, fagerberg1997tech, vanroy2018tech} all documented a positive relationship between R\&D and labor, whereas \citet{vanroy2018tech} found that this only holds in high- but not in low-tech industries. \citet{vainiomaki1999tech} found that high-technology sectors, measured by R\&D intensity, have the highest job creation rates. 
\citet{feldmann2013} provided only indirect support for the reinstatement effect: In the short term, increased innovation reflected in patent applications had a negative employment effect, but this effect diminished over time which indicates that employment was reinstated after a labor reducing technology shock. 

Three studies used different indicators of specific types of capital investment as proxies for technological change. 
\citet{ho2008invest} used a price index-based approach to capture quality improvements in equipment and found that these enhancements are associated with an increase in demand for non-production labor in US manufacturing. 
\citet{morrison1997} considered investments in specific high-tech equipment and \citet{raval2019micro} used the evolution of the capital stock as a technology measure. They all reported a positive relationship between labor demand and the capital indicators, suggesting a  complementarity between these factors. The results by \citet{morrison1997}, similar to \citet{ho2008invest}, suggest that this relation is particularly strong for non-production labor. 

Two studies with support for the reinstatement effect used an indirect automation risk-based approach \citep{gardberg2020digit, oesch2010}. Both confirmed at the country-level, and \citet{gardberg2020digit} additionally at the individual employee-level, that displacement is less likely in occupational groups with low automation risk. In contrast, employment has even increased in these jobs.

\citet{padalino1997employment,hoskins2000,madariaga2018employment,reijnders2018tech,jiang2019subst} 
are macro-level studies using indirect approaches to capture technological change, such as decomposition analyses and substitution elasticities. 
\citet{padalino1997employment,jiang2019subst} found a positive relationship between technological change and employment, while the other two found it only for certain sectors and jobs. However, also \citet{padalino1997employment} observed that the employment-GDP relationship is opposite in manufacturing, suggesting the simultaneous existence of replacement in manufacturing and reinstatement in non-manufacturing industries in recent decades.

Finally, four non-regression analyses in the sample of studies relying on the residual technology category found support for the reinstatement effect. 
\citet{reshef2013tech} relied on an empirically estimated simulation model calibrated on US data. The results confirm the hypothesis of skill-biased technological change in the US during 1963-2005 documenting an increasing demand for skilled labor at the expense of unskilled. 
For Canada, \citet{green2015} could not confirm the skill bias hypothesis in their detailed descriptive analysis. The analyses by \citet{buyst2018polar} and \citet{oesch2010} also suggest a bias of technological change being associated with increases in high paid occupations. Both studies also provide evidence for a rise in the demand for labor in the lowest skill group, but \citet{oesch2010} showed that this effect is heterogeneous across countries.

\subsubsection{Studies not supporting the reinstatement effect}
The two non-supporting studies are the studies by \citet{acemoglu2020robots, goaied2019}. While they do not support the reinstatement effect, they offer support for the replacement and real income effect. Both studies rely on regression analyses and report an overall negative impact of technology on net employment. \citet{acemoglu2019automation} studied the impact of robots on the regional employment to population ratio in 1990-2007 in the US. The authors found that robot-diffusion is associated with an increase in unemployment driven by employment losses in manufacturing and routine manual, blue collar occupations. The effects are larger for men than women. 
Additionaly, they reported positive productivity effects and increases in capital income. 

\citet{goaied2019} studied the long term impact of ICT-diffusion in 167 countries grouped into five regions using the number of mobile phone and internet users as a technology indicator. Distinguishing between long and short term effects, the authors reported negative short and long term employment effects. They reported a positive relationship between GDP and employment, but cannot attribute this to technology.

\subsubsection{Studies with ambiguous findings and indirect evidence}

Fifteen studies report ambiguous or only indirect evidence for the reinstatement effect. Six of them rely on ICT, six on innovation, one on TFP-style measures and two more looked, next to ICT, into the role of R\&D intensity (coded as other/indirect). 

The TFP-style study by \citet{dupaigne2009shock} estimated the macroeconomic employment effect of positive labor productivity shocks using a vector auto-regressive model. The authors found that the impact, whether positive or negative, is heterogeneous across countries and also depends on how technology shocks are measured.

Among the ICT studies, four of them rely on regression analyses  \citep{autor1998computers, fonseca2018polar, breemersch2019polar, omahony2021lshare}, one study conducted a simulation exercise \citep{charalampidis2019} and another one relies on descriptive statistics \citep{borland2017robot}. 
\citet{charalampidis2019} studied automation technology shocks (while not providing much detail about the measurement of these shocks) and analyzed their interaction with aggregate fluctuations of the labor share. While this study found that technology shocks explain a large share of the fluctuations, it remains inconclusive with regards to the longer term impact of these shocks. However, the author argued that labor reinstatement may be insufficient to offset job losses. 
The analysis by \citet{borland2017robot} relies on a detailed descriptive analysis of those industries in Australia that reported the largest changes in employment during the past few decades. They linked these observations to ICT-diffusion curves, but failed to identify any clear impact of technology on labor. The aggregate demand for labor was roughly constant over the considered period, but they also observed an increase in occupations that are intensive in non-routine tasks. 

Two of the four regression analyses in this sample investigated job polarization. \citet{fonseca2018polar} found support for technology as a driver of job polarization, which may be indicative for the reinstatement of certain types of jobs. \citet{breemersch2019polar} observed country-level employment growth and cross-industrial shifts, but failed to attribute this to technological change. \citet{autor1998computers} documented skill upgrading processes and higher wage premia, but did not show any clear impact on the demand for labor. The study by \citet{omahony2021lshare} investigated the relationship between ICT-diffusion and the labor share. They found a negative relationship between ICT capital investments and the labor share which may indicate lack of reinstatement. Interestingly, the authors also studied the impact of R\&D investments and observed the opposite effect.  

The six innovation papers \citep{evangelista2012inno, pellegrino2019rnd, piva2005innovation, piva2004change, vanreenen1997inno, kwon2015inno} report ambiguous or only weak support for the reinstatement effect. All of them are firm-level regression analyses and confirm a positive effect of product innovation on employment, but a negative effect for process innovation. Further, all of these studies, except for \citet{piva2004change, piva2005innovation}, simultaneously report ambiguous findings for the replacement effect.

\FloatBarrier
\subsection{Real income effect}
\label{sec:results_realincome}

\subsubsection{Overview, methods and technical issues}
\begin{table}[H] \centering \footnotesize 
\begin{threeparttable}
\caption{Studies by findings on real income}
\label{tab:realincome}
    \begin{tabular}{l*{4}{c}} \toprule
    & (1) & (2) & (3) & (4) \\
	\csname @@input\endcsname{inputs/stats_realinc_byfind}
	\bottomrule
	\end{tabular}
	\begin{tablenotes}[flushleft] 
	\scriptsize \item Notes: Columns (1)-(4) present the share and number (\#) of studies with empirical results that support, depend (on various characteristics, e.g. technology type, analysis level, etc.), are weak, and do not support the presence of a real income effect, respectively. The total \# of studies in the sample is 127 of which 33 examine the real income effect.
	\end{tablenotes}
\end{threeparttable}	
\end{table}

With thirty-three studies in total, the number of papers providing empirical insights on the real income effect is much smaller compared to the other two effects. Table \ref{tab:realincome} shows that among these, twenty-three papers found support, six report ambiguous (weak or depends) results, and four found no support.

The majority of studies investigate the impact of ICT (36\%), followed by robots (27\%), TFP-style (18\%) and other/indirect measures (15\%). Only two papers (6\%) study innovation (see column (1) from Table \ref{tab:tech_byfinding_realincome}). 

\begin{table}[H] \centering \footnotesize 
\begin{threeparttable}
\caption{Studies by findings on real income effect for each technology group considered}
\label{tab:tech_byfinding_realincome}
    \begin{tabular}{lc|*{4}{c}} \toprule
    & (1) & (2) & (3) & (4) & (5) \\
    & \multicolumn{1}{c}{}& \multicolumn{4}{c}{by finding} \\
    \cmidrule{3-6}	
	& \multicolumn{1}{c}{total}& \multicolumn{1}{c}{support}&\multicolumn{1}{c}{depends}&\multicolumn{1}{c}{weak}&\multicolumn{1}{c}{no support}\\ \midrule
	\csname @@input\endcsname{inputs/stats_tech_byfind_realinc}
	\bottomrule
	\end{tabular}
	\begin{tablenotes}[flushleft] 
	\scriptsize \item Notes: 	Column (1) presents the share and number (\#) of studies by the technology group considered in each row panel relative to the total number of studies examining the real income effect. The row-sum of shares in column (1) does not add up to one since there are studies considering more than one technology group. Columns (2)-(4) present the share and \# of studies by the set of findings reported on the real income effect for each type of technology considered in each row panel. For findings, a study is classified as `support' if it finds a significant empirical effect that supports the real income effect examined. `Depends' refers to the set of studies that find varying effects depending on the type of technology or the (sub)sample analyzed (e.g. type of sector or labor). Studies reporting effects that are negligible in terms of the magnitude were classified as `weak'. Studies were labeled as `no support' if they investigate the effect of interest, but documented insignificant or opposite effects. The technology types reported include ICT, robots, innovation, TFP-style and other types of technologies (e.g. indirectly measured through prices, automation risks), respectively.  Note that there is no overlap in findings reported, i.e. more than one primary finding in each study, and thus the shares across columns add up to one, up to rounding. The total \# of studies in the sample is 127 of which thirty-three examine the real income effect. 
	\end{tablenotes}
\end{threeparttable}	
\end{table}

The existence of the real income effect is empirically supported if studies found empirical support for at least one of the channels of technology-labor interactions (see Section \ref{sec:conceptual_framework}). The majority of papers (68\%) provides empirical evidence about the productivity channel, followed by income (29\%), and output (18\%) (see Table \ref{tab:realincome_byeffect}). Despite the high number of papers reporting results on the productivity effects of technological change, only 9\% are informative about the impact on prices, which is a relevant indicator to evaluate whether or not consumers benefit from the productivity gains of technological change. 

\begin{table}[H] \centering \footnotesize 
\begin{threeparttable}
\caption{Studies by type of real income effect examined}
\label{tab:realincome_byeffect}
    \begin{tabular}{l*{5}{c}} \toprule
    & (1) & (2) & (3) & (4) & (5) \\
	\csname @@input\endcsname{inputs/stats_grealinc_byfind}
	\bottomrule
	\end{tabular}
	\begin{tablenotes}[flushleft] 
	\scriptsize \item Notes: Columns (1)-(5) present the share and number (\#) of studies focusing on productivity, prices, income, output, and other real income effects, respectively. The total \# of studies in the sample is 127 of which 33 examine the real income effect. 
	\end{tablenotes}
\end{threeparttable}	
\end{table}

\subsubsection{Studies supporting the effect}


Among the real income supporting studies, ten investigated the impact of ICT \citep{autor2002computers, baddeley2008structural, chun2015it, oulton2002ict, strohmaier2016ict, vu2013ict, autor2015there, berman1994changes, blanas2019who, goaied2019}. Seven of them reported a positive impact of ICT on productivity, three found positive income effects, and two provide evidence of a positive relationship between output growth and labor demand. 
The three studies by \citet{autor2002computers, oulton2002ict, autor2015there} rely on descriptive analyses using industry-level case studies, macro-level growth accounting methods, and macroeconomic history. 
The other seven are regressions are macro- or meso-level, as well as one firm-level study by \citet{chun2015it}. 

Another six studies examined the impact of robots using Meso or regional-level regressions \citep{dekle2020robot, jung2020tech, acemoglu2019automation, acemoglu2020robots, blanas2019who, graetz2018robots}. They all provide evidence of the productivity increasing effect of robots. \citet{blanas2019who} also found a positive effect of robots on the wage bill of high-skill, old and middle aged men which may indicate an increase in the demand for certain types of jobs like engineers and managers. 

Five papers rely on TFP-style measures \citep{boyle2002trade, chen2014total, autor2018lshare, bessen2020autom, graham2000}. 
All of them are regressions, mostly at the macro-level except for the study by \citet{bessen2020autom} who performed a detailed analysis of the steel, auto and textile industry at a historical scale, and the region-industry-level study by \citet{graham2000}. Four of these studies reported positive productivity effects. \citet{boyle2002trade} and \citet{autor2018lshare} documented a positive impact on wage income, \citet{bessen2020autom} additionally found decreasing prices in response to labor productivity growth, and \citet{graham2000} confirmed that higher output is positively associated with the demand for labor. 

\citet{cirillo2017tech} examined the impact of innovation and found at the industry-level that an innovation-induced expansion of output and sales is positively associated with labor demand, especially for clerk and manual workers. 

The studies by \citet{fagerberg1997tech, padalino1997employment} used technology measures that fall into our residual category other/indirect. \citet{fagerberg1997tech} used R\&D as a technology indicator and \citet{padalino1997employment} used changing growth-employment elasticities as an indirect proxy of technological change. They both documented a positive relationship between aggregate output and employment.

\subsubsection{Studies not supporting this effect}

Only four studies did not find supporting evidence for the real income effect \citep{badescu2009impact,colombo2013ict,samaniego2006tech,dixon2020decline} not finding any positive impact of technology on productivity. \citet{dixon2020decline} provided additional explorations on income effects. 
\citet{badescu2009impact} and \citet{colombo2013ict} studied the impact of ICT-diffusion at the micro-level. \citet{dixon2020decline} used TFP-style measures, and \citet{samaniego2006tech} relied on indirect measures of technological change to  study the impact on macro-level TFP-growth. 
\citet{badescu2009impact} and \citet{colombo2013ict} could not confirm any significant relationship between productivity and ICT-diffusion, while \citet{dixon2020decline} and \citet{samaniego2006tech} reported even negative effects. \citet{dixon2020decline} additionally found that the impact of technology on income is negative. 
Based on an empirically estimated general equilibrium simulations, \citet{samaniego2006tech} studied the impact of productivity shocks. The author argued that the negative productivity effect of a new technology can be explained by the incompatibility of an existing technology with a new one. However, \citet{samaniego2006tech} also argued that this may be only temporary.

\subsubsection{Studies with ambiguous findings and indirect evidence}

Six studies in our sample show ambiguous findings or only indirect evidence for the real income effect \citep{camina2020autom,fu2021,blien2021occup,compagnucci2019robotization,cozzarin2016inno,oesch2010}. 
The studies by \citet{camina2020autom, fu2021, compagnucci2019robotization} studied the impact of robot-diffusion. \citet{camina2020autom} showed that whether robots are productivity-enhancing depends on the exact type of robot-based technology; for example, some technologies like data-driven control can even exhibit a negative effect. 
\citet{fu2021} reported cross-country differences: While robots show a positive effect on labor productivity in developed economies, no significant effects are found for developing countries. 
\citet{compagnucci2019robotization} found ambiguous results: They observed a negative effect of robots on wages, but a negative one on prices. 

\citet{blien2021occup, oesch2010} relied on other/indirect measures. In particular, \citet{blien2021occup} looked at the routine-intensity of occupations and evaluated its interaction with the income of certain occupations. They observed that the effects on employees' income are heterogeneous across occupations: Employees in jobs with high routine-intensity experience income losses after a job layoff. The authors also showed that routine-intensity was a less significant predictor of income losses during the 1980s compared to more recent periods. In a descriptive macro-level analysis, \citet{oesch2010} used the changes in the wage growth of various occupations that vary in their task content. They examined the evolution of wages for different occupations but found only weak effects.

Finally, \citet{cozzarin2016inno} studied the impact of innovation on wages and productivity in manufacturing and only found weak aggregate effects. In tendency, \citet{cozzarin2016inno} observed a positive association between process innovation, wages and productivity, but a negative one for product innovation and productivity.

\section{Discussion}
\label{sec:discussion}

What is the net employment effect of technological change? It depends on which of the labor-saving or the labor-creating effects of technological change dominate. We systematically collected evidence for technology-induced labor replacement and creation by disentangling three mechanisms of labor-technology interactions: 1) direct replacement; 2) direct reinstatement; and 3) indirect real income effects. The first mechanism is labor-saving, while the latter two labor-creating. 
In this section, we discuss the extent to which our analysis enables us to draw conclusions about the \emph{net} employment effects of technology. 

\subsection{Evidence for net employment effect}
In total, we found at least as many studies supporting the labor-creating effects (n=64 for the reinstatement and n=23 for the real income) as studies supporting the replacement effect (n=69). 

One concern is the existence of a publication bias, whereby studies with significant support for an effect are more likely to be published \citep{ugur2019inno}. 
We found that within the three subsets of studies reporting evidence for one of the effects, the shares of studies supporting the respective effect are roughly comparable, i.e. 67\% for the replacement, 81\% for the reinstatement, and 70\% for the real income effect. 
These numbers suggest that if there was a publication bias, the bias would be roughly equal across all effects. 

\begin{table}[H] \centering \footnotesize 
\begin{threeparttable}
\caption{Studies by findings on net employment}
\label{tab:totalemp}
    \begin{tabular}{l*{4}{c}} \toprule
    & (1) & (2) & (3) & (4) \\
	\csname @@input\endcsname{inputs/stats_total_byfind}
	\bottomrule
	\end{tabular}
	\begin{tablenotes}[flushleft] 
	\scriptsize \item Notes: Columns (1)-(4) present the share and number (\#) of studies with empirical results reporting positive, depends (on various characteristics, e.g. technology type, analysis level, etc.), weak, and negative effects on net employment, respectively. The total \# of studies is 127.
	\end{tablenotes}
\end{threeparttable}	
\end{table}

The heterogeneity in the design of studies does not allow us to systematically compare the sizes of the effects reported. However, some studies provide direct insights on the net employment effect. Specifically, Table \ref{tab:totalemp} shows that a larger share finds support for an overall positive (29\%) than a negative employment effect (18\%). 
Overall, our findings strongly suggest that technological progress has not resulted in a negative net employment effect in the past decades. 
In the next section, we summarize and contextualize the insights gained from our analysis about the net employment effect for each technology. 

\begin{table}[H] \centering \footnotesize 
\begin{threeparttable}
\caption{Each type of technology by findings on net employment effect}
\label{tab:tech_byfinding_total}
    \begin{tabular}{lc|*{4}{c}} \toprule
    & (1) & (2) & (3) & (4) & (5) \\
    & \multicolumn{1}{c}{}& \multicolumn{4}{c}{by finding} \\
    \cmidrule{3-6}	
	& \multicolumn{1}{c}{total}& \multicolumn{1}{c}{positive}&\multicolumn{1}{c}{depends}&\multicolumn{1}{c}{weak}&\multicolumn{1}{c}{negative}\\ \midrule
	\csname @@input\endcsname{inputs/stats_tech_byfind_total}
	\bottomrule
	\end{tabular}
	\begin{tablenotes}[flushleft] 
	\scriptsize \item Notes: Column (1) presents the share and number (\#) of studies by the technology group considered in each row panel relative to the total number of studies examining the net employment effect. The row-sum of shares in column (1) do not add up to one since there are studies considering more than one technology group. 
	Columns (2)-(4) present the share and \# of studies by the set of findings reported on the net employment effect for each type of technology considered in each row panel. For findings, a study is classified as `positive' if it finds a significant positive empirical effect that supports the net employment effect examined. `Depends' refers to the set of studies that find varying effects depending on the type of technology or the (sub)sample analyzed (e.g. type of sector or labor). Studies reporting effects that are negligible in terms of the magnitude were classified as `weak'. Studies were labeled as `negative' if they investigate the effect of interest, but documented negative effects. The technology types reported include ICT, robots, innovation, TFP-style and Other types of technologies (e.g. indirectly measured through prices, automation risks), respectively.  Note that there is no overlap in findings reported, i.e. more than one primary finding in each study, and thus the shares across columns add up to one, up to rounding. The total \# of studies is 127 of which seventy-five examine the net employment effect.
	\end{tablenotes}
\end{threeparttable}	
\end{table}

\subsubsection{Net employment effects of ICT}
The most ambiguous results are found in ICT studies. 
For the replacement effect, we observed simultaneously a high number supporting (n=23) and a high number not (n=9) or only weakly (n=5) supporting the effect. 
For the reinstatement effect, we found twenty-three supporting it. These studies confirm that ICT reinstates labor in new ICT-related occupations. The findings also suggest that the dissemination of ICT is related to an employment increase for high-skill workers and non-routine cognitive labor suggesting that ICT complements workers in the performance of non-routine tasks. 
We also found a sample of studies reporting an ICT-related employment growth in the service sector. 
For the real income effect, we identified seven studies reporting a positive effect of ICT on productivity, and a few other studies reporting positive income effects and a positive relationship between output growth and labor. 
Most studies support both the Replacement and Reinstatement effect at the same time. 

In line with these observations, eleven studies (42\%, Table \ref{tab:tech_byfinding_total}) that report net employment effects find that the impact depends on other factors, with the most important being the type of labor. 
We also found a larger share reporting a positive (27\%) rather than a negative (8\%) effect on net employment. 

We conclude that ICT does not appear to induce an overall net negative effect, but our results show that the jobs reinstated qualitatively differ from the jobs replaced. These findings have important policy and management implications. Although our results suggest that the employment perspectives have developed more favorably for high-skill workers whose skills complement ICTs, investing in the right type of skills will also benefit (future) workers in low- and middle-skill occupations. 

Despite the fact that many occupations in the middle - and also some in the lower end -  of the skill distribution are intensive in tasks that can be performed by ICTs, many jobs in these segments will continue to require a changing set of skills due to increasingly digitalized work environments \citep{autor2015there, levy2010technology}. 
Think for example of automotive technicians who use computer systems to diagnose cars. These workers are required to have basic digital skills to access  computer systems, but also the ability to interpret information in digital environments. The need to invest in the development of digital skills is also stressed by the 2020 European Skills Agenda \citep{european2020european}. One of the European Skill Agenda's objective is to significantly increase the share of adults with a basic level of digital skills. Moreover, various EU initiatives have been introduced to promote the development of digital skills in vocational education and training institutions and systems \citep{european2022vocational}. 

Our results also indicate a shift of work from production to services within and across sectors. Hence, the skills demanded in the digitalized economy are not only \textit{prima facie} digital skills, but also social and emotional skills are required to perform well in service tasks \citep{autor2013polar}. The growing importance of social skills in the labor market is also reflected in the increasing wage returns for these skills \citep{deming2017growing}.

\subsubsection{Net employment effects of robots}
The highest relative support for the replacement effect comes from studies analyzing the impact of robots.
Out of the studies investigating the replacement effect of robots, thirteen (87\%, Table \ref{tab:tech_byfinding_replacement}) found support for this mechanism. 
Nine studies support the reinstatement effect of robots and six studies support the real income effect finding positive productivity and income effects. 
A large share of the reinstatement studies also reports net employment effects, but none of them finds clear evidence for a negative effect of robots on net employment.
Also when looking at the full sample of robot studies (Table \ref{tab:tech_byfinding_total}), we found that a larger share provides support for a positive net employment effect (36\%) rather than a negative (21\%). 
A substantial share of robot studies exploring the total employment effect report negligible effects. 

Overall, our findings suggest that the labor-saving impact of robots tends to be offset by robot-driven reinstatement of labor. 
These results confirm a recent turn in the academic debate about the impact of robots and automation on jobs, suggesting that the fear of a jobless future \citep{smith2017automation, mokyr2015history} may be exaggerated and lack an empirical base \citep{economist2022robots, aghion2022effects}. This finding has important political implications; for example, it undermines the rationale of a robot tax which has been put forward as a means to cope with automation-induced unemployment.

Although robots and ICTs have replaced workers in the performance of routine tasks, robots differ from ICTs as they are particularly suitable to perform manual tasks. Moreover, in comparison to ICTs, robots are mainly considered as a pure automation technology with relatively clear-cut effects on labor demand while the effects of ICT are more ambiguous. 

\subsubsection{Net employment effects of innovation}
The least support for the replacement effect is found by innovation studies. Only three studies support the real income effect (25\%, Table \ref{tab:tech_byfinding_replacement}). 
The innovation studies showing support for the replacement effect found that the labor-saving effect is mostly driven by process innovation, but overall, the available evidence is insufficient to conclude that process innovation is labor-saving. 
Those studies that did not support the replacement effect or reported ambiguous evidence suggest a positive effect of product innovation on employment. 
However, a high number of studies remain inconclusive and found no significant or only weak effects of product, organizational or process innovation on employment. 

For the reinstatement effect, we identified eight (57\%, Table \ref{tab:tech_byfinding_reinstatement}) studies supporting the effect. Another four studies (29\%) indicate that the reinstatement effect depends on the type of innovation. In general, studies seem to find a positive employment effect in the case of product innovation and a negative or ambiguous effect of process innovation. 

Thirteen innovation studies provided insights on the net employment effect, but none of them found a negative effect. Six studies suggest that the impact is positive, while seven report ambiguous results.

In line with other reviews \citep{vivarelli2014innovation, ugur2017litreview, calvino2018innovation}, our findings suggest that the distinction between process and product innovation is essential when speaking about the employment effects of innovation. 
It is not surprising that process innovation can be labor-saving as it is by definition an innovation aimed to make production processes more efficient, which is mostly equivalent with input-saving. However, process innovation is not necessarily labor-saving for three reasons: (1) Process innovation can also save other inputs rather than labor; (2) efficiency improvements may lead to lower prices and an expansion of output which can be associated with the creation of new jobs; and (3) process innovation can coincide with the introduction of technologies that complement human labor, creating labor in industries where new machines are produced.

It is also not surprising that product innovation tends to be labor-creating. The successful launch of a product innovation is associated with the acquisition of new markets which may lead to an expansion of economic activity of a firm, and thus with the creation of new jobs. 


\subsubsection{Net employment effects of TFP-style technology}
Among the TFP-style studies, we found strong support for the simultaneous destruction of production labor, mostly in manufacturing, and creation of new non-production labor, mostly in services. 
We identified thirteen studies (76\%, Table \ref{tab:tech_byfinding_replacement}) that support the replacement effect. Many of them showed labor demand shifts, particularly from production to non-production labor within and across industries, and from low- to high-skill workers. 
For the reinstatement effect, we found ten studies (91\%, Table \ref{tab:tech_byfinding_reinstatement}) supporting it by documenting employment shifts from manufacturing to services and employment increases for skilled non-production labor. 
Moreover, one reinstatement study showed that TFP shocks in upstream industries are associated with employment increases in downstream industries. 
These observations support the idea that technological change leads to structural change with a reallocation of economic activity down the value chain, i.e. from more primary and secondary industries towards increasingly processed sectors and services \citep{kruger2008productivity, syrquin1988patterns}. 

We also identified five studies supporting the real income effect that report rising wage incomes and decreasing prices. One study also confirmed a positive relationship between income and employment. The number of TFP-style studies finding support for the replacement effect is roughly balanced compared to the number of studies finding support for one of the two labor-creating mechanisms. 

Among the TFP-style papers, only thirteen provided insights on net employment effects, among which only one suggests a positive impact, five suggest a negative impact, and the remaining found that the net impact is weak or depends. 
The theory of structural change suggests a reallocation of labor from 
high to low productivity growth industries with increasing capital intensity and automation as major drivers of productivity growth \citep{syrquin1988patterns}. Theoretical models of structural change further suggest that the elasticity of demand is decisive for the sufficient reinstatement of labor \citep{ngai2007structural} which we aimed to capture through the assessment of the real income effect. 
Our results for the net employment effect are rather inconclusive, but suggest that the net impact is more negative than positive. This finding indicates that the reinstatement of labor in more downstream and service-related industries may be insufficient to offset the labor-saving impact. 
However, as a critical note, it should be flagged that historically productivity and welfare increases coincided with a reduction of weekly working hours, which may be seen as a very desirable effect \citep{bick2018hours}.

\subsubsection{Net employment effects of other/indirect technology}
The support for the labor-saving and labor-creating effects from studies relying on other/indirect measures is roughly balanced. 
We found twenty-two studies supporting the replacement effect (73\%, Table \ref{tab:tech_byfinding_replacement}) and twenty studies (91\%, Table \ref{tab:tech_byfinding_reinstatement}) supporting the reinstatement effect. Another two studies (40\%, Table \ref{tab:tech_byfinding_realincome}) provide support for the Real income effect. {Furthermore, we find that three studies do not report any of the effects discussed above, but provide results on net employment.\footnote{These are macro-level studies by \cite{samaniego2008tech, balleer2012new, karabarbounis2013lshare}, with the former reporting positive net employment effects, while the remaining studies suggest the presence of negative net employment effects.}}

A first set of studies analyzed the relationship between automation risk indicators and employment. Studies supporting the replacement effect largely found negative employment effects in occupations with high automation risks, while studies that support the reinstatement effect reported employment increases in low automation risk jobs. 

A second set of studies used capital and high-tech equipment investments as measures of technological change. Also these studies found support for both the replacement and reinstatement effect. 
The same holds for studies looking at the impact of R\&D expenditures and patents. 
One study on the real income effect also reported a positive relation between R\&D, aggregate output and employment. 
\textcolor{red}{}

Studies on the replacement effect found that especially low-skill labor is negatively affected. However, we also found a few descriptive studies providing evidence that the demand for low-skill labor can also be positively affected, especially in service jobs. 
Overall, the results suggest a bias of technological change that leads to a reallocation of labor towards non-production labor and high-tech industries, which is often, but not necessarily, high-skill. These findings provide both support for the skill-biased technological change hypothesis \citep{acemoglu2002technical} as well as for the routine-biased technological change hypothesis that predicts a polarization of the labor market \citep{autor2003skill}.

Furthermore, we found that a number of indirect measures of technology only provide evidence for one of the three employment mechanisms. One replacement effect study reports a negative association between GDP growth and employment in manufacturing, while a number of decomposition-like and substitution elasticity-based studies found support for the reinstatement effect. 
Finally, one study on the real income effect uses growth-employment elasticities as an indirect measure for technology and reports a positive relationship between aggregate output and employment. Again, we conclude that indirect measures of technological change provide a roughly equal amount of evidence for the labor-saving effect and for the labor-creating effect. 

In total, twenty-nine studies in the category other/indirect also investigated the net employment effect. The share of papers reporting an overall positive net employment effect (28\%) is larger than the share of papers documenting an overall negative effect (21\%). A substantial share of papers report that the overall employment effect depends on other factors (41\%). Hence, the set of studies using alternative measures do not support the idea that technological change has resulted in a net destruction of jobs.

These indirect measures relying on automation risks have a number of disadvantages. For instance, the primary error engendered by automation risk indicators is that they implicitly assume that tasks that can be performed by machines will automatically substitute for human labor. 
Moreover, occupation-level automation risk indicators are likely to overestimate the share of automatable jobs as they disregard task heterogeneity within occupations as well as the adaptability of jobs in response to technological change \citep{arntz2017}.

\section{Conclusions}
\label{sec:conclusion}
This study systematically reviewed the available evidence on the impact of technological change on employment. Overall, we found that a substantially larger number of studies provide evidence for the labor-creating impact of technological change than for the labor-displacing impact. Several studies providing support for the labor-creating impact of technology report positive effects on productivity and prices, income, and final demand or output. Through these channels, technological change is expected to indirectly increase the demand for labor. However, only a few studies measure the actual employment implications of these mechanisms. 

Although we are careful in concluding that, if anything, technological change has a positive net effect on employment, we do conclude that the replacement effect is likely to be more than offset by the labor-creating effect of technology. Hence, there does not appear to be an empirical foundation for the fear of technology-driven unemployment. We also investigated whether different types of technology have a differential effect on total employment. For almost all technology measures (i.e. robots, ICT, TFP-style, other/indirect measures) we found a comparable number of studies that support the labor-displacing as well as labor-creating effect of technology. Only for innovation measures, the empirical evidence suggests that product innovation is mostly labor-creating, while the evidence concerning process innovation remains inconclusive. 

Despite the fact that we found no evidence for a negative net employment effect in quantitative terms, the qualitative impact of technological change on employment cannot be neglected. As our systematic review has pointed out, different types of technology have adverse effects on predominantly low-skill and production labor, manufacturing jobs, and workers performing routine tasks. Given the considerable labor-saving potential of technology, reskilling or retraining workers whose jobs are susceptible to automation is essential. However, not only individuals whose jobs have disappeared due to automation would benefit from (re-)training. Research shows that technological change also has induced within-occupational changes in skill requirements. In fact, according to estimates of the OECD \citep{arntz2017}, the share of jobs that will face changes in the task content due to automation is higher than the share of jobs that is of high risk of being automated. Hence, upskilling is also of utmost importance for workers whose jobs are not directly threatened by technological change. Furthermore, some displaced workers may not be able to make the transition to new jobs after experiencing job loss. These workers might have to rely on social support systems.
  
Our systematic review is subject to a number of limitations. First, empirical studies can only cover the impact of technologies that are already available today, but the scope of tasks that may be automated in the near future continuously expands \citep{brynjolfsson2014second, frey2017future}. Hence, empirical evidence on the impact of artificial intelligence, quantum computing, virtual reality, biotechnology, nanotechnology, renewable energy, and other emerging technologies that will soon impact our economy remains limited. In fact, none of the studies in our review assessed the impact of this new wave of technological innovation. To that end, it is unclear to which extent our findings can be extrapolated into the future. Therefore, real-time monitoring and ongoing research is needed to more fully understand the emerging impact of the introduction of new technologies on the future of work \citep{baldwin2019globotics}. 

Second, our study also faces an inherent methodological challenge. Generally, it should be noted that it is not surprising to observe high support rates for each of the effects as studies with insignificant results are rarely published. Hence, these results are subject to a reporting bias, similar to the publication selection bias discussed in \citet{ugur2019inno}. This appears to be a more general problem of published empirical studies which also exists in other disciplines such as medicine, psychology, and experimental economics research. In these fields, pre-registrations were introduced with a detailed protocol of the empirical approach that has to be approved before the experimental study is conducted. While similar procedures are more difficult to establish in statistical analyses that often need to be adapted during the workflow, it should be kept in mind that the results reported are likely to suffer from a reporting bias in favor of the analyzed theory. Nevertheless, there is no reason to assume that the size of the bias differs across the three employment effects that we analyze in our study.

\newpage
\printbibliography

\newpage
\FloatBarrier
\appendix
\renewcommand{\appendixname}{Appendix}
\renewcommand{\thesection}{\Alph{section}} \setcounter{section}{0}
\renewcommand{\thefigure}{\Alph{section}.\arabic{figure}} \setcounter{figure}{0}
\renewcommand{\thetable}{\Alph{section}.\arabic{table}} \setcounter{table}{0}
\renewcommand{\theequation}{\Alph{section}.\arabic{table}} \setcounter{equation}{0}

\newpage
\begin{center}
    \textbf{\LARGE{}Supplementary Online Appendix}{}
\end{center}

\section{Search strategy and selection procedure of studies} \label{appendix:searchstrategy}

We closely followed the PRISMA 2020 guidelines to ensure the quality of the systematic search process \citep{page2021litreview}. A scoping review was used to identify relevant search terms in widely cited studies. 
Subsequently, a computerized search was performed using the search term that appeared either in the title, abstract, or list of keywords of studies, namely: `technolog*' combined with `labo\$r' or `employment'. 
Specifically, the search strings used in Web of Science included: 
\begin{itemize}
    \item Query: (TS=((technolog*) AND (labo\$r OR *employment)) AND WC=(ECONOMICS OR MANAGEMENT OR BUSINESS OR BUSINESS, FINANCE OR SOCIOLOGY OR INDUSTRIAL RELATIONS LABOR OR DEVELOPMENT STUDIES OR SOCIAL SCIENCES INTERDISCIPLINARY OR HISTORY OR SOCIAL ISSUES OR URBAN STUDIES OR GEOGRAPHY)  AND LA=(English) AND DT=(Article OR Book Chapter OR Early Access OR Proceedings Paper OR Review))
    \item Timespan: 1988-01-01 to 2021-04-21 (Publication Date)
    \item Web of Science Index: Social Sciences Citation Index (SSCI)
\end{itemize}

The search was conducted in the Web of Science (WoS) Core Collection database. We opted for WoS because of its proven suitability as a principal search engine for systematic reviews \citep{gusenbauer2020litreview}. As we are interested in the economic impact of technological change, we only included records from the Social Sciences Citation Index (SSCI) of WoS and limited our search to the categories: Economics; Management; Business; Business, Finance; Sociology; Industrial Relations and Labor; Development Studies; Social Sciences, Interdisciplinary; History; Social Issues; Urban Studies; and Geography. The search was restricted to the most relevant document types and included articles, book chapters, early access documents, proceeding studies, and reviews, and documents in English language. The search covered articles published during the time period 1988-April 2021.

The initial search in WoS resulted in 8,748 studies published between 1 January 1988 and 21 April 2021.\footnote{See Appendix Figure~\ref{fig:selectionprocedure} for an overview of the selection process of relevant studies.} 
Given the large search outcome, six independent researchers were involved in  screening the relevant records to ensure the timely and unbiased completion of the process. 
The procedure was structured in such a way that each researcher evaluated 1/6 of the retrieved records in a first round, i.e. approximately 1,450 per researcher. In a second round, the six sets of records were rotated such that a second researcher re-assessed the previously screened records. To avoid bias, the procedure ensured that in the second round, two separate researchers were assigned to the set of records screened by the original reviewer in the first round. To illustrate, if researcher A screened articles 1 to 1,450 in the first round, then in the second round, researcher B screened articles 1 to 725 and researcher C screened articles 726 to 1,450. The first and second rounds of screening were conducted independently such that the assessment of the first reviewer was unknown to the second set of reviewers.

Based on the title and abstract, every researcher screened each record and assessed whether it is considered for this review as: relevant (``yes''); potentially relevant (``maybe''); or irrelevant (``no''). In this first step, studies were considered relevant if the independent variable is related to technology and the dependent variable is related to (un)employment (inclusion criterion 1). 
In a second step, records that were considered relevant by both researchers (``yes''/ ``yes'') or relevant by one and potentially relevant by the other (``yes''/``maybe'') were kept. Vice versa, records that coded with (``no''/``no'') and (``no''/``maybe'') were excluded. 
Mutual uncertainties (``maybe''/ ``maybe'') and strong disagreements (``yes''/``no'') were subjected to further reviewing by a randomly assigned third reviewer, who made the final decision whether to in- or exclude the article. 

In step 3, the remaining 252 studies were assigned to three researchers, i.e. the authors of this paper, who each independently assessed 1/3 of the remaining studies against the following inclusion criteria: (1) examines the effect of technological change on employment; 
(2) makes a significant empirical contribution, i.e. excluding purely theoretical studies; and (3) covers at least one developed country and the period after the 1980s.\footnote{Here, countries are defined as developed if classified as high-income by the \cite{worldbank2021countrygroup}.} 
Regarding (1), we included studies relying on direct and indirect measures of technological change such direct measures of technology adoption and use, but also indirect indicators reflecting structural changes in the production technology. For employment, we considered direct measures of labor demand such as employment, hours worked, and employment rates, but also measures that provide insights on the dynamics at the labor market such as wages, wage bill, employment shares of different skill or occupational groups, and substitution elasticities. 
Finally, we did not impose any specific restrictions on the level of analyses. Therefore, our final selection of studies investigates the impact of technological progress at various levels of aggregation, i.e. at the macro (e.g. country), meso (e.g. sectors, industries), micro (e.g. firms, individuals), regional (e.g. regions, states, cities), and other (e.g. skills, occupations) levels of data aggregation. 

Step 4 of the systematic search led to the inclusion of 127 studies. These studies were coded along a scheme that was iteratively refined throughout the process. 
We recorded the countries studied, the period covered, the outcome variable(s), and extracted descriptive information about the empirical operationalization of technological change, the measurement of the employment effects, the insights that can be gained about the three effects (replacement, reinstatement, and real income), and, if applicable, information about the net employment effect of technologies. 
We also extracted information about the level of analysis (e.g. country, region, firm, employee, occupation), data sources, sample size, methodology, potential heterogeneity, if applicable, and bibliometric information about the author, publication year and outlet.

\begin{landscape}
\begin{figure}[ht] \centering
\caption{Overview of the selection procedure of studies}
\label{fig:selectionprocedure}
    \vspace{-0.3cm}
    \includegraphics[height=1\textwidth]{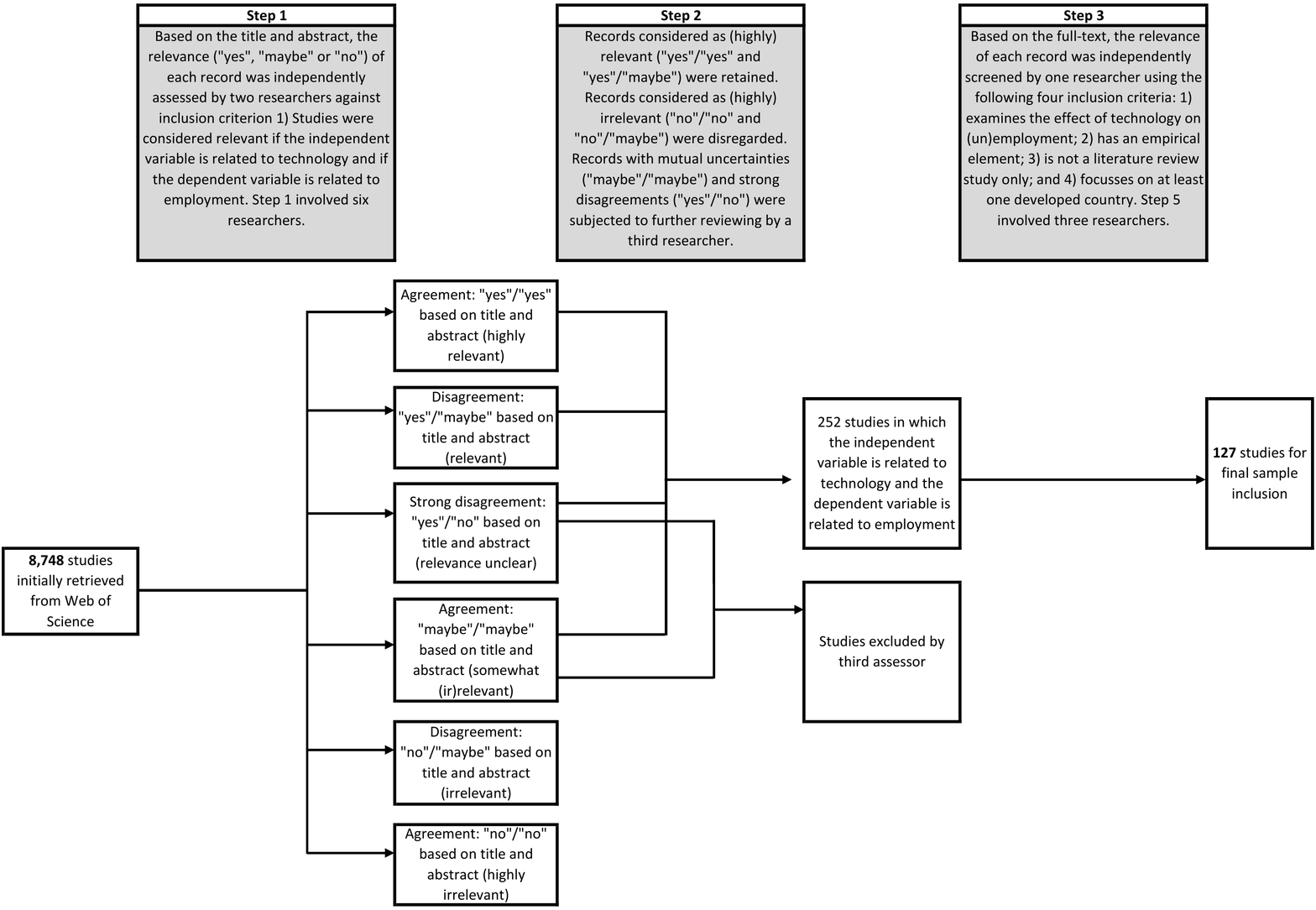}
    \vspace{-1cm}
    \begin{fignote} 
    Source: Author's illustration.\\
    \end{fignote}
\end{figure}
\end{landscape}

\newpage
\section{Additional Tables and Figures}

\begin{table}[H] \centering \footnotesize 
\begin{threeparttable}
\caption{Studies by the type of empirical methodology used in the analysis for any and each effect explored}
\label{tab:effect_bymethod}
    \begin{tabular}{l*{3}{c}} \toprule
    & (1) & (2) & (3) \\
    & \multicolumn{1}{c}{Descriptive}&\multicolumn{1}{c}{Regression}&\multicolumn{1}{c}{Simulation}\\ \midrule
	\csname @@input\endcsname{inputs/stats_effect_bymethod}
	\bottomrule
	\end{tabular}
	\begin{tablenotes}[flushleft] 
	\scriptsize \item Notes: Columns (1)-(3) present the share and number (\#) of studies by the primary type of empirical methodology used in each study to identify the effect(s) of interest. `Descriptive' refers to studies using descriptive statistics and conceptual analyses that link macro-level stylized facts to empirically reported technology-trends at the micro-level. `Regression' refers to any regression-based analysis or other quantitative inferential methods with empirical foundation. `Simulation' captures simulation methods, e.g. DSGE. These statistics are reported in each row panel within any, replacement, reinstatement, real income, and net employment effect reported, respectively. Note that there is no overlap in methods reported, i.e. more than one primary method used in each study, and thus the shares across columns add up to one, up to rounding.
	\end{tablenotes}
\end{threeparttable}	
\end{table}

\begin{table}[H] \centering \footnotesize 
\begin{threeparttable}
\caption{Findings for each effect, including na/nr}
\label{tab:effect_byfinding_incl_nanr}
    \begin{tabular}{l*{5}{c}} \toprule
    & (1) & (2) & (3) & (4) & (5) \\
	&\multicolumn{5}{c}{Replacement}\\
	\csname @@input\endcsname{inputs/stats_replace_byfind_incl_nanr} \midrule
	&\multicolumn{5}{c}{Reinstatement}\\
	\csname @@input\endcsname{inputs/stats_reinst_byfind_incl_nanr}  \midrule
	&\multicolumn{5}{c}{Real income}\\
	\csname @@input\endcsname{inputs/stats_realinc_byfind_incl_nanr} \midrule
	&\multicolumn{5}{c}{Total employment}\\
	\csname @@input\endcsname{inputs/stats_total_byfind_incl_nanr} 
	\bottomrule
	\end{tabular}
	\begin{tablenotes}[flushleft] 
	\scriptsize \item Notes: Columns (1)-(4) present the share and number (\#) of studies with empirical results that support, depend (on various characteristics, e.g. type of technology or level of analysis), are weak, and do not support the presence of the effect considered in each row panel. Column (5) presents the share and number of studies out of the total sample of studies which do not examine the effect of interest. The row row-panels present results for the replacement, reinstatement, real income and total employment effect. The total \# of studies is 127.
	\end{tablenotes}
\end{threeparttable}	
\end{table}

\begin{table}[H] \centering \footnotesize 
\begin{threeparttable}
\caption{Reinstatement effect by level of analysis}
\label{tab:reinstatement_bylevel}
    \begin{tabular}{l*{4}{c}} \toprule
    & (1) & (2) & (3) & (4) \\
    & \multicolumn{1}{c}{macro}&\multicolumn{1}{c}{meso}&\multicolumn{1}{c}{micro}&\multicolumn{1}{c}{regional}\\ \midrule
	\csname @@input\endcsname{inputs/stats_reinst_bylevel}
	\bottomrule
	\end{tabular}
	\begin{tablenotes}[flushleft] 
	\scriptsize \item Notes: Columns (1)-(4) present the share and number (\#) of studies employing macro, meso, micro, and regional level of analysis, respectively when reporting any type of finding on the reinstatement effect.
	\end{tablenotes}
\end{threeparttable}	
\end{table}

\begin{table}[H] \centering \footnotesize 
\begin{adjustbox}{width=1\textwidth}
\begin{threeparttable}
\caption{List of studies}
\label{tab:studies1}
    \csname @@input\endcsname{inputs/studies1}
	\begin{tablenotes}[flushleft] 
	\scriptsize \item Note: Continued on next page
	\end{tablenotes}
\end{threeparttable}	
\end{adjustbox}
\end{table}
\begin{table}[H] \centering \footnotesize 
\begin{adjustbox}{width=1\textwidth}
\begin{threeparttable}
\caption{List of studies – continued from previous page}
\label{tab:studies2}
    \csname @@input\endcsname{inputs/studies2}
	\begin{tablenotes}[flushleft] 
	\scriptsize \item Note: Continued on next page
	\end{tablenotes}
\end{threeparttable}	
\end{adjustbox}
\end{table}
\begin{table}[H] \centering \footnotesize 
\begin{adjustbox}{width=1\textwidth}
\begin{threeparttable}
\caption{List of studies – continued from previous page}
\label{tab:studies3}
    \csname @@input\endcsname{inputs/studies3}
	\begin{tablenotes}[flushleft] 
	\scriptsize \item Notes: This table presents descriptive information about the 127 studies used in this review after a systematic search of the literature on technology and jobs. This includes information about: the author and publication year; various classifications of interest based on the technology group, level of analysis (e.g. country, region, firm, employee, occupation), and empirical methods used in each study; and the findings about net employment and focal effects, i.e. replacement, reinstatement, and real income, of technologies. Finally, for the real income effect we also present the relevant variable of interest. See Section~\ref{sec:methods} for more details on the coding scheme.
	\end{tablenotes}
\end{threeparttable}	
\end{adjustbox}
\end{table}

\end{document}Concerns that automation and technological advancement will make human labor obsolete is not a recent phenomenon. Already during the first Industrial Revolution, the adoption of power looms and mechanical knitting frames gave rise to the Luddite movement. The Luddites protested against disruptive technology by destroying textile machinery out of fear of job loss and skill obsolescence. The idea that